\documentclass{aa}
\include{psfig}
\usepackage{graphics}

\newcommand{\rms}{{\rm rms}}
\newcommand{\cd}{\cdot}
\newcommand{\cdd}{\!\cd\!}
\newcommand{\ti}{\times}
\newcommand{\tii}{\!\ti\!}
\newcommand{\ds}{\displaystyle}
\newcommand{\lb}{\langle}
\newcommand{\rb}{\rangle}
\newcommand{\na}{\nabla}
\newcommand{\pa}{\partial}

\newcommand{\mb}[1]{\mbox{#1}}

\begin{document}
\title{Magnetoconvection and dynamo coefficients:}
\subtitle{Dependence of the $\alpha$ effect on rotation and magnetic field}
\titlerunning{$\alpha$ effect: dependence on rotation and magnetic field}
\author{M. Ossendrijver\inst{1} \and M. Stix\inst{1}
        \and A. Brandenburg\inst{2,}\inst{3}}
 \institute{Kiepenheuer-Institut f\"ur Sonnenphysik,
              Sch\"oneckstra{\ss}e 6, 79104 Freiburg, Germany
 \and Department of Mathematics, University of Newcastle upon Tyne, NE1 7RU, UK
 \and Nordita, Blegdamsvej 17, DK-2100 Copenhagen \O, Denmark}
\date{Received; accepted}

\abstract{
We present numerical simulations of three-dimensional compressible magnetoconvection
in a rotating rectangular box that represents a section of the solar convection zone. 
The box contains a convectively unstable layer, surrounded by stably 
stratified layers with overshooting convection. The magnetic Reynolds number,
Rm, is chosen subcritical, thus excluding spontaneous growth of the magnetic
field through dynamo action, and the magnetic energy is maintained by 
introducing a constant magnetic field into the box, once convection has attained
a statistically stationary state. Under the influence of the Coriolis force,
the advection of the magnetic field results in 
a non-vanishing contribution to the mean electric
field, given by $\lb\vec{u}\ti\vec{b}\rb$. From this electric field, we calculate
the {\em $\alpha$-effect}, separately for the stably and the unstably stratified
layers, by averaging over time and over suitably defined volumes. From the
variation of $\alpha$ we derive an error estimate, and the dependence of
$\alpha$ on rotation and magnetic field strength is studied. Evidence is
found for rotational quenching of the {\em vertical} $\alpha$ effect, and
for a monotonic increase of the {\em horizontal} $\alpha$ effect with
increasing rotation. For $\mb{Rm}\approx 30$, our results for both 
vertical and horizontal $\alpha$ effect are consistent with magnetic
quenching by a factor $[1+\mb{Rm}\,(B_0/B_{\rm eq})^2]^{-1}$. The signs
of the small-scale current helicity and of the vertical component of
$\alpha$ are found to be opposite to those for isotropic turbulence.
}

\maketitle

\section{Introduction}
Magnetic fields are observed on a wide variety of cosmical bodies, among which are planets, stars and
galaxies. With the exception of a few types of objects whose magnetic fields are thought to be 
frozen-in relic fields, cosmical magnetic fields are attributed to dynamo action.
Dynamo theory concerns the generation of magnetic fields in electrically conducting fluids. 
In stars and planets, dynamo action is the result of an interplay between convection and rotation.
The present simulations focus on dynamo action in late-type stars,
which are characterized by an outer convective zone on top of a stably 
stratified radiative interior. Many late-type stars are magnetically
active, and some exhibit magnetic cycles, as does the Sun.

A successful solar model was first proposed by Parker (1955) who recognized
that shear and helicity cooperate in such a way that the mean magnetic field
oscillates in a migratory manner; later Yoshimura (1975) has shown that the
field generally propagates along the surfaces of constant angular velocity.
Since then, magnetic cycles and butterfly diagrams have been produced by
mean-field models using spherical
geometry (Steenbeck \& Krause 1969), with depth-dependent magnetic
diffusivity (Roberts \& Stix 1972), and with the character of an 
interface wave at the base of the convection zone (Parker 1993,
Charbonneau \& MacGregor 1997). As an alternative to the helical effect of
convection (the $\alpha$~effect), the systematic tilt of flux tubes that
become unstable and erupt to the solar surface has been employed
as a regenerative agent for the mean poloidal field (Leighton 1969; 
Durney 1995, 1997; Schmitt et al. 1996; Dikpati \& Charbonneau 1999), with
equal success regarding the existence of Sun-like cycles and butterfly
diagrams. In all cases, however, the regenerative terms in the mean-field
equations were based either on approximations such as first-order 
smoothing, or on the limited available knowledge about the behavior of
magnetic flux tubes in the convection zone; the review of Stix (2001)
summarizes some of the current problems.

In principle, the problem can be attacked by direct numerical integration
of the equations of magnetohydrodynamics (MHD) for the star's convection
zone (for a review, see e.g. Nordlund et al. \cite{nordlund94}). 
But severe problems are encountered.
Firstly, a stellar convection zone is a highly turbulent plasma
($\mb{Re}>10^{12}$), so that the kinetic energy spectrum encompasses a very
wide range of length scales. If these were all to be resolved, a prohibitive 
number of mesh points would be required. Secondly, since the magnetic
Reynolds number is also large ($\mb{Rm}>10^8$), a similar argument holds
for the length scales of the magnetic field. For these reasons
three-dimensional MHD simulations often have been restricted to a 
rectangular box that represents only a small section of a stellar 
convection zone. In the present work we do this, too. Thirdly, the Prandtl
number, the ratio of the kinematic viscosity to the radiative diffusivity,
is very small in the solar convection zone ($\mb{Pr}\approx 10^{-4}-10^{-7}$), 
so that the flow can vary on much smaller scales than the temperature.
This is known to have consequences for the topology and temporal
behavior of convective turbulence
(Cattaneo et al. \cite{cattaneo91}; Brandenburg et al. \cite{brandenburg96}; 
Brummell et al. \cite{brummell96}), but 
in the present simulations, we ignore such effects, and set $\mb{Pr}=1$.
Fourthly, the Mach number, $\mb{Ma}=u/c_{\rm s}$, decreases with depth in the 
solar convection zone from about $0.1$ in the photosphere to less than
$0.001$ near the base. The corresponding high sound speed in the lower 
convection zone  necessitates the use of very small time steps in order to
fulfill the CFL condition for compressible hydrodynamics 
($\Delta t<\Delta x/c_{\rm s}$), while it is often desirable to continue
simulations for at least several convective turnover times. One possibility
to circumvent this problem is to adopt the anelastic approximation
whereby sound waves are excluded, so that larger time steps can be 
used (Ogura \& Charney \cite{ogura62}; Lantz \& Fan \cite{lantz99}).
Here we maintain full compressibility, and attain a mean Mach number 
of the order $0.1$; in a subsequent paper we shall study effects of a
smaller Mach number. 

Brandenburg et al. (\cite{brandenburg90}) extended a 3D hydrodynamical
code to the case of magnetoconvection, including the effect of rotation.
Subsequently, a spontaneous dynamo instability was observed in the
simulations (Nordlund et al. \cite{nordlund92}, Brandenburg et al. 
\cite{brandenburg96}). Recent simulations of isotropically forced helical
turbulence (Brandenburg \cite{brandenburg00}) have verified the
existence of large scale dynamo action, and it was possible to identify
this as the result of an $\alpha$ effect (in the sense of a
{\it non-local} inverse cascade). However, the large scale
field generated possesses magnetic helicity and, since for closed or
periodic boundaries the magnetic helicity can change only resistively,
the growth of the large scale field is slowed down as the magnetic
Reynolds number increases.  This translates inevitably to a 
magnetic-Reynolds-number dependent $\alpha$ effect and
turbulent magnetic diffusivity, as suggested by Vainshtein
\& Cattaneo (\cite{vainshtein92}). The argument of 
Vainshtein \& Cattaneo is however only phenomenological, not based 
upon the fundamental concept of magnetic helicity conservation, and
hence their conclusion that strong large-scale fields are
impossible is not borne out by the simulations of
Brandenburg (\cite{brandenburg00}). Furthermore, in that paper it 
was shown that the $\alpha$ obtained by imposing a magnetic field is 
indeed a reasonable approximation to the $\alpha$ that results 
naturally even if no field is imposed.

In the present simulations we do not calculate box dynamos, but rather
concentrate on the dynamo coefficients that occur in mean-field theory.
Therefore, all simulations are performed with an imposed magnetic field,
and the magnetic Reynolds number is chosen subcritical, so that the
magnetic field is not self-sustained. We think that progress in stellar
dynamo theory can be made through a combination of exact MHD simulation
and mean-field dynamo theory. Furthermore, the existence of 
systematic, large-scale magnetic fields and stellar cycles with periods
far in excess of convective time scales suggests that a mean-field
description is possible in some form. Also, we argue that mean-field
theory (in the wide sense that includes the above-mentioned models
such as Leighton's) is the only current model that reproduces
large-scale magnetic fields and cycles, including such an outstanding
feature as the solar butterfly diagram, even though the diverse
approximations made for calculating the dynamo coefficients may not
be valid under stellar conditions. 
The case of self-excited dynamo action would of course be very
interesting too, especially in view of the question of whether
$\alpha$-quenching depends on the magnetic Reynolds number, as suggested
by Vainshtein \& Cattaneo (\cite{vainshtein92}). However, we postpone this
until a later study, partly because the measurement of the dynamo
coefficients in the presence of a large-scale field
that is different from the imposed one is less straightforward.

The present investigation is limited to the regime where 
the coefficients are determined directly by the flow, and 
we do not address the question of whether the large-scale magnetic
field of late-type stars is in fact generated by such an ordinary 
$\alpha$ effect in the convection zone, or by a magnetic instability 
in an underlying stably stratified layer
(e.g., Brandenburg \& Schmitt \cite{brandenburg98}).
We do include a stably stratified 
layer with overshooting convection, but its main purpose here is to 
provide realistic conditions for the flow in the unstable region.
In most runs, the strength of the imposed field is set to a value which 
amounts to typically $2\%$ of the equipartition field with respect
to the kinetic energy density. During the subsequent evolution
the field strength also remains small compared to the equipartition value. 
Furthermore we shall explore the influence of the imposed magnetic field
by increasing its strength up to values somewhat in excess of the 
equipartition value.
Therefore the results may have relevance for the solar convection zone, 
where the magnetic field is no stronger than the equipartition value. 
The $\alpha$ effect produced in the galactic
gas by supernova explosions has been calculated by Ziegler et al.
(\cite{ziegler96}), in a similar spirit as in the present work. 

The relevance of the transport coefficients for mean-field dynamo theory
becomes clear from the equation for the mean magnetic field,
\begin{equation}
\frac{\pa\lb\vec{B}\rb}{\pa t}=\na\tii\big[\lb\vec{U}\rb\tii\lb\vec{B}\rb+
   \lb\vec{u}\tii\vec{b}\rb-\eta\na\tii\lb\vec{B}\rb\big]\,,
\end{equation}
where $\lb\vec{B}\rb$ is the mean magnetic field, 
$\lb\vec{U}\rb$ is the mean flow, $\vec{u}=\vec{U}-\lb\vec{U}\rb$
is the fluctuating component of the flow, and $\lb\vec{u}\ti\vec{b}\rb$
is a contribution to the mean electric field, sometimes referred to as the 
{\em electromotive force}.
The dynamo coefficients appear in an expansion of this mean electric
field in terms of spatial derivatives of the mean magnetic field. 
In  general  they can be represented as  kernels of an
integral equation (Brandenburg \& Sokoloff \cite{brandenburg99}).
In the simplest case, the coefficients are treated as local tensors, which leads to
\begin{equation}
\langle\vec{u}\ti\vec{b}\rangle_i=\alpha_{ij}\langle B_j\rangle+
    \beta_{ijk} \na_j\langle B_k\rangle+\cdot\cdot\cdot\,.
\end{equation}
The first term in the expansion is the $\alpha$ effect. The alpha tensor
is a pseudo tensor that exists only in non mirror-symmetric flows, as
they occur in stellar convection. For the solar dynamo, the main 
significance of the $\alpha$ effect lies in generating the poloidal mean
magnetic field, which is achieved predominantly by $\alpha_{\phi\phi}$.
Toroidal fields are generated by the strong differential rotation 
that exists in the {\em tachocline} near the base of the solar convection
zone. In other solar-type stars and fully convective stars though, differential rotation could be 
weak (K\"uker \& Stix \cite{kueker00}), so that the dynamo becomes of the $\alpha^2$-type. In that case, 
other components of the alpha tensor, such as $\alpha_{rr}$, are 
important for maintaining the toroidal magnetic field.

The remainder of the paper is structured as follows. After introducing the model and
the equations, we focus on the dependence of $\alpha$ on rotation. 
In the following section, effects of the strength and orientation of the imposed field 
are studied. In the final section, a discussion of the results is presented.

\section{The Model}
\begin{figure}
\centerline{\psfig{file=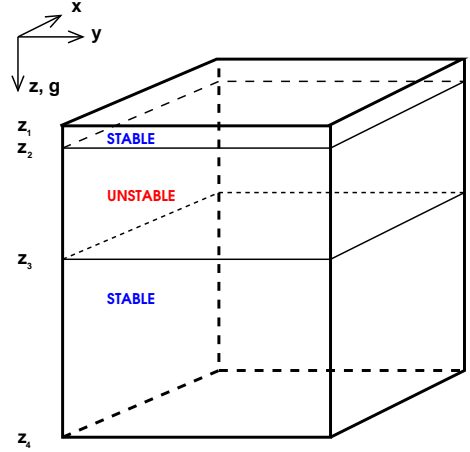,width=6cm}}
\caption{Geometry of the simulation domain. The layer boundaries are at
         $z_1=-0.15$, $z_2=0$, $z_3=1$, and $z_4=2.85$, except for run 10
         that has no stable regions, so that $z_1=z_2=0$ and $z_3=z_4=1$.
         The horizontal extent is 4.}
\label{fig1}
\end{figure}

The model used in the simulations is the same as that of Brandenburg et al.
(\cite{brandenburg96}). The simulation domain consists of a rectangular box
which is defined on a Cartesian grid, with $x$ and $y$ denoting the two
horizontal coordinates (corresponding to latitude and longitude in spherical
coordinates), and $z$ denoting depth, respectively (Fig.~\ref{fig1}).
In most cases, convectively stable layers are included  below and above the
unstable layer. The upper layer (region 1) is stabilized by a cooling term
in the energy equation, which leads to an almost isothermal, highly stable
stratification. The lower stable layer (region 3) represents an overshoot
zone, whose thickness is chosen such that overshooting plumes do not
reach the lower impenetrable boundary. The box rotates about an axis 
$\vec{\Omega}$, that makes an angle $\theta$ with the $z$-axis. Hence
$\theta=0$ corresponds to a situation where the box is located at the south 
pole of the Sun. In the present paper we treat only this special case.

The governing equations are those describing magnetic induction, mass 
continuity, and the balance of momentum and energy:

\begin{eqnarray}
\frac{\pa\vec{B}}{\pa t}&=&\na\tii (\vec{U}\tii\vec{B})+\eta\na^2\vec{B}\,,\\
\frac{d\rho}{dt}&=&-\rho\na\cdd\vec{U}\,,\\
\rho\frac{d\vec{U}}{dt}&=&-\na p+\rho\vec{g}+\vec{J}\tii\vec{B}
      -2\rho\,\vec{\Omega}\tii\vec{U}+2\nu\na\cdd\rho\tens{S}\,,\\ 
\rho\frac{de}{dt}&=&-p\na\cdd\vec{U}+\na\cdd\kappa\na e
      +2\nu\!\rho\tens{S}^2+\mu_0\eta J^2+Q\,,
\end{eqnarray}
where $\vec{J}=\nabla\times\vec{B}/\mu_0$ is the current density,
$\eta$ is the magnetic diffusivity, $\nu$ is the kinematic viscosity, 
$\gamma=C_p/C_V$ is the adiabatic index, and $\kappa$ is the radiative 
conductivity. In all simulations, $\eta$ and $\nu$ are taken constant, and 
$\kappa$ is a prescribed function of depth. The stress tensor $\tens{S}$ is 
given by
\begin{equation}
{\sf S}_{ij}=\frac{1}{2}\Bigg(\frac{\pa U_i}{\pa x_j}+\frac{\pa U_j}{\pa x_i}
\Bigg)-\frac{1}{3}\delta_{ij}\na\cdd\vec{U}\,,
\end{equation}
and $\tens{S}^2$ stands for $\sum_{ij}{\sf S}_{ij}^2$. 
The equation of state is that of an ideal gas, i.e.,
\begin{equation}
p=(\gamma-1)\rho e\,.
\end{equation}
The term $Q$, given by 
\begin{equation}
Q=-\sigma_0 f(z)\,(e-e_1)\,, \label{coolingterm}
\end{equation}
represents cooling or heating, depending on whether the internal energy 
density exceeds $e_1$ or falls below it, respectively; $\sigma_0$ represents 
the cooling/heating rate. The depth-dependent function $f$ ensures that $Q$ 
is non-vanishing only in region 1, and $f(z_2)=0$. The inclusion of this 
term leads to the presence of a highly stable, thin overshoot layer, thereby
providing a realistic upper boundary condition for the convection zone. 
In the horizontal directions, periodic boundary conditions are imposed. 
The upper and lower boundaries of the domain are impenetrable and 
stress-free, and the horizontal components of the magnetic 
field variation are set to zero. At the lower boundary, the energy flux 
is prescribed, and at the top boundary the internal energy 
(which is proportional to the temperature) is fixed.
\begin{equation}
\begin{array}{ll}
z=z_1,z_4\hspace{1cm}&\left\{\begin{array}{ll}
 \ds\frac{\pa u_x}{\pa z}=\frac{\pa u_y}{\pa z}=u_z=0\,,\\[4mm]
 B_x=B_{0x}\,,\\[3mm]
 B_y=B_{0y}\,,\end{array}\right.\\[1.4cm]
z=z_1 & e=e_1\,,\\[4mm] 
z=z_4  & \ds\frac{de}{dz}={\Big(}\frac{de}{dz}{\Big)}_4\,.
\end{array}\label{boundy}
\end{equation}
In the actual runs we set $B_{0y}=0$; $B_{0x}$ is either 0 or the 
imposed horizontal field.

All quantities are made dimensionless by setting
\begin{equation}
d=\rho_0=g=\mu_0=C_p=1\,,
\end{equation}
where $\rho_0$ is the initial density at a depth $z=z_3$, the bottom of
the unstable layer. Thus length is expressed in terms of $d=z_3-z_2$, the
thickness of the convectively unstable zone (region 2). It follows that time
is measured in terms of $\sqrt{d/g}$, velocity in terms of $\sqrt{gd}$, the
magnetic field strength in terms of $\sqrt{\mu_0\rho_0 gd}$, and entropy
in terms of $C_p$.

On several occasions we shall consider the energy balance, 
which is governed by a conservation law, 
\begin{equation}
\frac{\pa (\rho e_{\rm tot})}{\pa t}=-\sum_i\na\cdd\vec{F}_i+Q\,,
\end{equation}
where $e_{\rm tot}= e+ |\vec{u}|^2/2+|\vec{B}|^2/2\mu_0\rho$ is the total 
specific energy, and the fluxes, $\vec{F}_i$, are given by
\begin{eqnarray}
\vec{F}_{\rm conv}&=&\gamma\rho e\vec{u}\,,  \\
\vec{F}_{\rm kin}&=&\rho |\vec{u}|^2\vec{u}/2\,,  \\
\vec{F}_{\rm rad}&=&-\kappa\na e\,,  \\
\vec{F}_{\rm visc}&=&-2\nu\!\rho\,\tens{S}:\vec{u}\,,  \\
\vec{F}_{\rm em}&=& \vec{E}\tii\vec{B}/\mu_0\,.
\end{eqnarray}
Apart from the geometry, boundary conditions, and initial conditions, solutions are governed by
the following 10 independent dimensionless parameters. The Prandtl number and the magnetic Prandtl 
number are defined as 
\begin{equation}
\mb{Pr}=\nu/\chi_0\,, \hspace{1cm} \mb{Pm}=\nu/\eta\,, \label{pr}
\end{equation}
where $\chi_0=\kappa_2/\gamma\rho_0$ denotes a reference value of the thermometric (radiative) 
diffusivity for region 2. The parameter $\xi_0$ determines the pressure scale height at the top 
of the box, $\xi_0=H_p(z_1)/d$:
\begin{equation}
\xi_0=(\gamma-1)\,e_1/gd\,.
\end{equation}
The initial thermal structure of the (maximally) three regions is characterized by the radiative 
temperature gradients,
\begin{equation}
\na_i={\Big(}\frac{d\ln T}{d\ln p}{\Big)}_i\hspace{2cm}(i=1,2,3)\,.
\end{equation}
In addition, one often employs the polytropic index, $m=(1-\na)/\na$. The adiabatic temperature 
gradient is given by $\na_{\rm ad}=(\gamma-1)/\gamma$. A measure for instability is provided by 
the superadiabaticity, $\delta_i=\na_i-\na_{\rm ad}$, which is positive in an unstably stratified 
medium. The Rayleigh number is defined as 
\begin{equation}
\mb{Ra}=\frac{d^4 g\delta_2}{\nu\chi_0 H_{ph}}\,, \label{ra}
\end{equation}
where $H_{ph}=\xi_0 d+0.5 d/(m_2+1)$ is the pressure scale height in the middle of the 
unstable layer, as can be shown using the hydrostatic equilibrium (\ref{e0}). The 
parameters $\delta_2$ and $H_{ph}$ refer to the unperturbed stratification, i.e. that before 
the onset of convection. The Rayleigh number is a measure of the strength of convection 
compared to that of viscous and thermal (radiative) dissipation, as can be seen by writing 
$\mb{Ra}=t_{\rm visc}t_{\rm rad}/t^2_{\rm conv}$, where $t_{\rm visc}=d^2/\nu$,
$t_{\rm rad}=d^2/\chi_0$, and $t_{\rm conv}^2=H_{ph}/g\delta_2$. Alternatively, one may 
express $\mb{Ra}$ in terms of the entropy gradient, $ds/dz=C_p\,\delta/H_p$. According to the Schwarzschild criterion, a positive value for $\delta$,
i.e. $\mb{Ra}>0$, signifies instability. In reality, Ra must exceed a
finite threshold value for convection to set in. It should be noted that
the {\em local} value of the Rayleigh number, 
$d^4 g\delta/(\nu\chi H_p)$, varies with depth within the unstable layer, because $\chi$, 
$H_p$ and, as a result of convection also $\delta$, are $z$-dependent. Typically, the local 
value is several times smaller than $\mb{Ra}$, mainly because $\delta$ can be strongly 
reduced by convection. The Taylor number, 
\begin{equation}
\mb{Ta}=(2\Omega d^2/\nu)^2\,,
\end{equation}
measures the importance of rotation relative to viscous dissipation. Finally, $\sigma_0$
represents the rate at which internal energy is lost from the upper stable layer, and 
$\theta$ is the angle between the rotation vector and the $z$-axis.

All other parameters are secondary. The Coriolis number, or inverse Rossby number, measures 
the importance of the Coriolis force and is defined as 
\begin{equation}
\mb{Co}=2\Omega \tau\,,
\end{equation}
where $\tau=\ell/u_{\rms}$ is the turnover time. The correlation length,
$\ell$, is taken to be $d$ in the unstable region. The Chandrasekhar number,
\begin{equation}
\mb{Ch}=\frac{\mu_0 B_0^2 d^2}{4\pi\rho_0\nu\eta},
\end{equation}
measures the strength of the imposed magnetic field. In the initial state, the $z$-component of the 
radiative energy flux, $F_{{\rm rad},z}=-\kappa de/dz$, is assumed to be constant throughout the domain. This
determines the radiative conductivities in the three regions
according to $\kappa_i/\kappa_2=(m_i+1)/(m_2+1)$. In fact, $\kappa$ is turned into a smooth function of depth 
by allowing it to change continuously across thin intermediate layers between the three regions. 
An approximate initial stratification, unperturbed by convection, is calculated on the assumption of 
hydrostatic equilibrium, and this is done iteratively until the condition $\rho(z=z_3)=\rho_0$ is satisfied. 
It is also assumed that region 1 is cooled efficiently enough to become isothermal. 
The result is then basically a smoothed version of 
\begin{eqnarray}
e&=&\left\{\begin{array}{ll}
\ds e(z_1)\,\Big[1+\frac{z-z_1}{(m_1+1)H_p(z_1)}\Big] &(\mb{i}),\\[4mm]
\ds e(z_2)\,\Big[1+\frac{z-z_2}{(m_2+1)H_p(z_2)}\Big]\hspace{1cm} & 
                                                      (\mb{ii}),\\[4mm]
\ds e(z_3)\,\Big[1+\frac{z-z_3}{(m_3+1)H_p(z_3)}\Big] & (\mb{iii}),\\
\end{array}\right.\label{e0}\\[5mm]
\rho&=&\left\{\begin{array}{ll}
\ds \rho(z_1)\,\Big[1+\frac{z-z_1}{(m_1+1)H_p(z_1)}\Big]^{m_1}\hspace{0.5cm}& (\mb{i}),\\[4mm]
\ds \rho(z_2)\,\Big[1+\frac{z-z_2}{(m_2+1)H_p(z_2)}\Big]^{m_2} & (\mb{ii}),\\[4mm]
\ds \rho(z_3)\,\Big[1+\frac{z-z_3}{(m_3+1)H_p(z_3)}\Big]^{m_3} & (\mb{iii}),\\
 \end{array}\right.\label{rho0}
\end{eqnarray} 
where $(\mb{i})\cd\cd\cd (\mb{iii})$ stand for the three regions, $z_1\leq z< z_2$, 
$z_2\leq z< z_3$, and $z_3\leq z\leq z_4$, respectively. The actual initial stratification 
in region 2 is calculated numerically using the mixing-length formalism of convection. The
advantage of this approach is that it reduces the amount of
time required for relaxation to a fully convective state. 
The internal energy density at the top equals 
$e_1=\xi_0 m_{\rm ad}gd$. Using (\ref{e0}), it is easily shown that
$(de/dz)_4=m_{\rm ad}g/(m_3+1)$. The radiative, kinetic and magnetic 
diffusivities follow from Eqs.~(\ref{pr}) and (\ref{ra}). 
Reynolds numbers are defined as 
\begin{equation}
\mb{Re}=u_{\rm rms}d/\nu\,, \hspace{1cm} \mb{Rm}=u_{\rm rms}d/\eta\,,
\end{equation}
where $u_{\rm rms}$ is the rms velocity defined in a suitable way (e.g., by averaging over time and 
over a partial volume of the box).
\begin{table*}[htb]
\caption{Run parameters. For the geometry, see Fig. \protect{\ref{fig1}}. The 
         imposed magnetic field is $\vec{B}_0=B_0\vec{e}_z$, or $B_0\vec{e}_x$,
         depending on whether $\alpha_{\rm V}$ or $\alpha_{\rm H}$ is
         determined. In all runs, the polytropic indices are given by $m_1=0$,
         $m_2=1$ and $m_3=3$, the mesh size is $50^3$ (except for run 10,
         where it is $30^3$); $B_0=10^{-3}$,
         $\mb{Ra}=5\cd 10^4$, $\mb{Pr}=\mb{Pm}=1$ (i.e. $\mb{Rm}=\mb{Re}$),
         and $\xi_0=0.2$. It follows that $\nu=0.00211$, and $\mb{Ch}=0.018$.
         The parameters $\mb{Ma}$, $\mb{Re}$ and $\mb{Co}$ are averages over
         time and over the unstable layer; $T$ is the time interval over
         which the averaging is performed.}
\begin{center}
\begin{tabular}{ccccccccccc} \hline\noalign{\smallskip}
run                    & 1      & 2      & 3      & 4      & 5      & 6      & 7      & 8      &  9      & 10   \\
\noalign{\smallskip}\hline\noalign{\smallskip}
Ta                     & $2$     & $10$    & $30$    & $100$   & $300$   & $2000$  & $10000$ & $20000$ & $50000$ & $10000$ \\ 
$T$ ($\alpha_{\rm V}$) & $206$   & $822$   & $206$   & $841$   & $698$   & $545$   & $940$   & $362$   & $376$ & 554  \\
$T$ ($\alpha_{\rm H}$) &  --     &  --     &  --     & $564$   & $564$   & $456$   & $508$   & $346$   & $290$ & 521  \\
Ma                     & $0.089$ & $0.088$ & $0.088$ & $0.084$ & $0.091$ & $0.089$ & $0.080$ & $0.070$ & $0.056$  & 0.10  \\
Re                     & $34$    & $34$    & $34$    & $34$    & $35$    & $34$    & $30$    & $30$    & $20$   &  38 \\
\mb{Co}                & $0.042$ & $0.093$ & $0.16$  & $0.30$  & $0.50$  & $1.3$   & $3.3$   & $5.4$   & $11$  &  2.6 \\
\noalign{\smallskip}\hline
\end{tabular}
\end{center}
\label{paramtable}
\end{table*}

We employ a finite difference scheme, according to which spatial derivatives
are calculated with 6th-order accuracy (Lele \cite{lele92}). Time-stepping is
done using a third-order Hyman predictor-corrector method. Table \ref{paramtable} gives a list
of the parameters used for a first series of runs in which the influence of
rotation is investigated by varying the Taylor number.

\begin{figure}[htb]
\centerline{\psfig{file=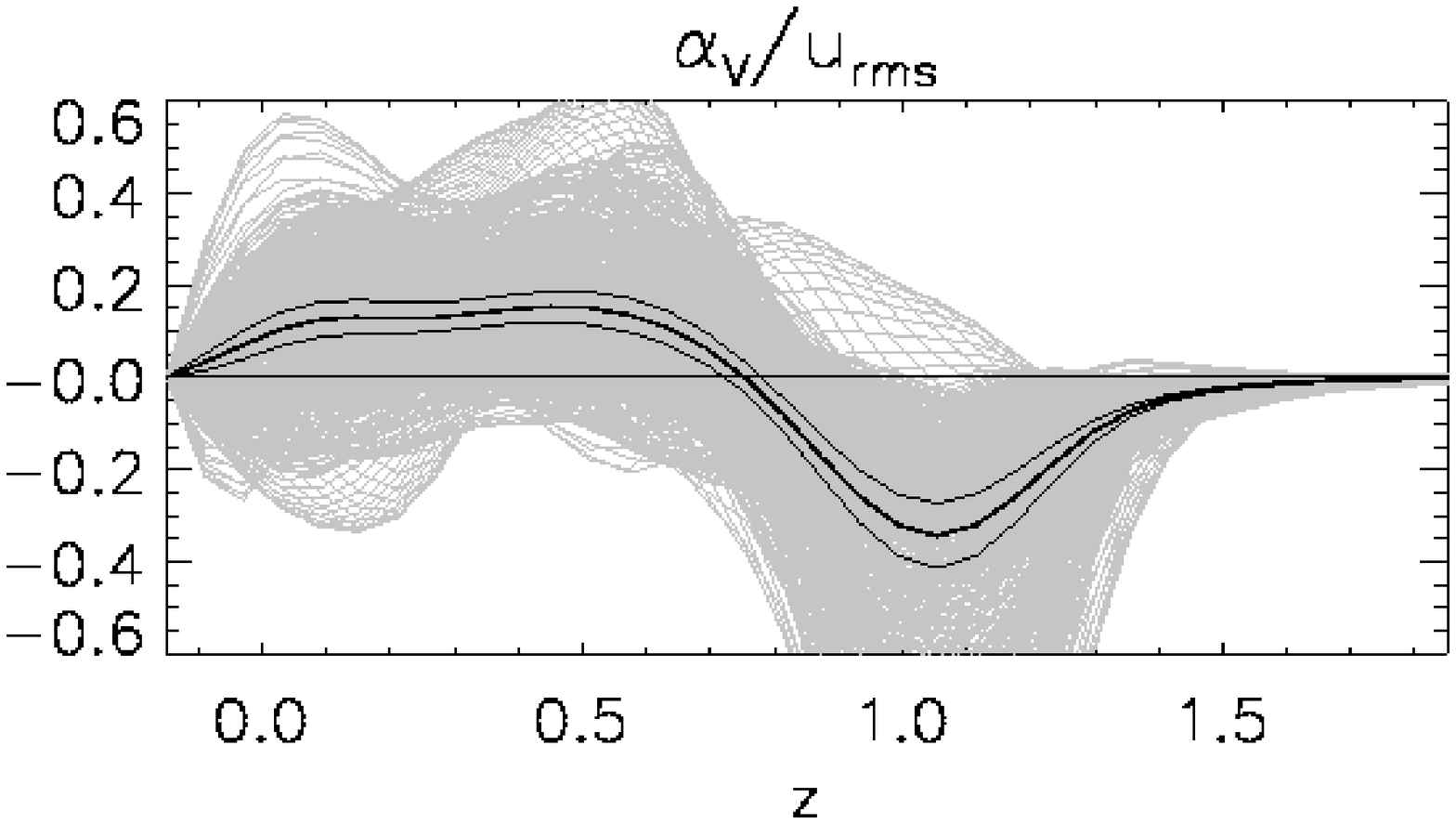,width=8.5cm}}
\centerline{\psfig{file=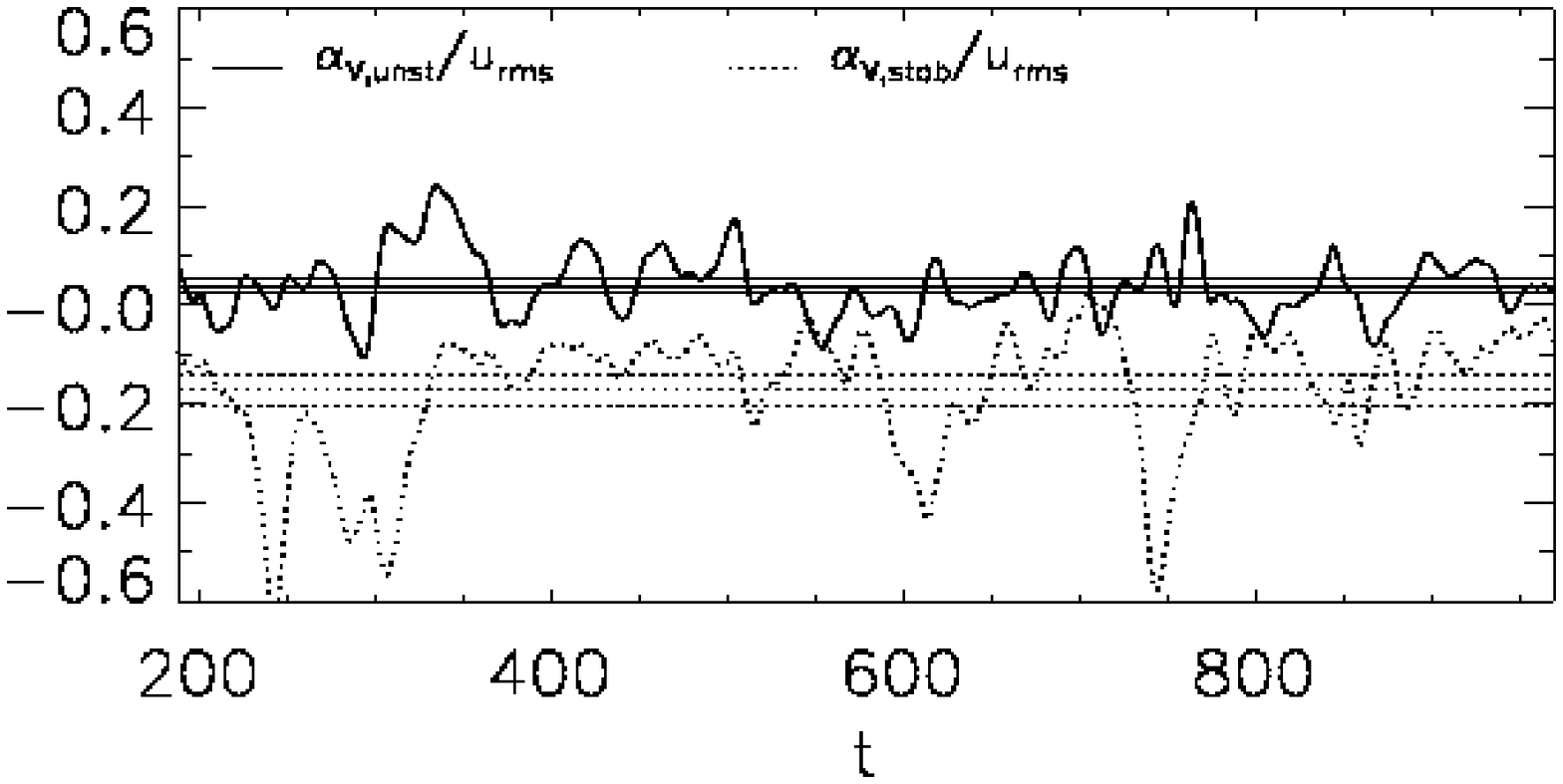,width=8.5cm}}
\caption{Averaging procedure (run 6). {\em Top}\/: at each time step, 
$\alpha_{\rm V}$ is calculated as a function of depth by averaging over the
horizontal coordinates (grey curves). The thick drawn curve is the
time average, and the surrounding thin drawn curves indicate the error in
the mean. {\em Bottom}\/: $\alpha_{\rm V}$ is calculated separately for two
partial volumes, 
namely $\alpha_{\rm V\!, unst}$ for the depth range $-0.15<z<1$,
and $\alpha_{\rm V\!, stab}$ for the range $1<z<1.5$. 
The error in the time averages is determined on the basis of the number of
coherence times covered by the simulation. All calculations are for the
solar south pole.}                                         \label{fig2}
\end{figure}

\section{The $\alpha$ effect}
Essentially, the $\alpha$ effect amounts to a proportionality between the
mean electromotive force, $\lb\vec{u}\tii\vec{b}\rb$, and the mean magnetic field.
Magnetoconvection in a rotating, stratified medium provides the necessary
anisotropies for a non-zero $\alpha$. In the present case, with the rotation
axis chosen parallel to the direction of gravity, the most elementary,
non-isotropic form of the expression for the $\alpha$ effect is one that
distinguishes vertical and horizontal components: 
\begin{equation}
\lb\vec{u}\tii\vec{b}\rb=\alpha_{\rm V}\lb \vec{B}_z\rb
                        +\alpha_{\rm H}\lb\vec{B}_{\rm H}\rb\,.
\end{equation}
The components $\alpha_{\rm H}=\alpha_{xx}=\alpha_{yy}$ and
$\alpha_{\rm V}=\alpha_{zz}$
correspond to $\alpha_{\phi\phi}$ and $\alpha_{rr}$ in spherical coordinates,
respectively. We recall that $\alpha_{\phi\phi}$ plays the dominant role in
the generation of the poloidal magnetic field. The two components of 
$\alpha$ are extracted from the simulations by constructing $\alpha_{\rm V}=
\lb\vec{u}\tii\vec{b}\rb\cdd\lb\vec{B}_z\rb/|\lb \vec{B}_z\rb|^2$ and 
$\alpha_{\rm H}=
\lb\vec{u}\tii\vec{b}\rb\cdd\lb\vec{B}_{\rm H}\rb/|\lb\vec{B}_{\rm H}\rb|^2$.
In practice, $\lb\vec{u}\tii\vec{b}\rb$ in these expressions is replaced by 
$\lb\vec{U}\tii\vec{B}\rb$, because the difference, 
$\lb\vec{U}\rb\tii\lb\vec{B}\rb$, was found to be negligible in every case. 
Depending on which component of $\alpha$, i.e. $\alpha_{\rm V}$ or 
$\alpha_{\rm H}$, is to be determined, a magnetic field is imposed in the 
$z$-direction or in the $x$-direction, respectively.

\begin{figure*}[htb]
\centerline{
\psfig{file=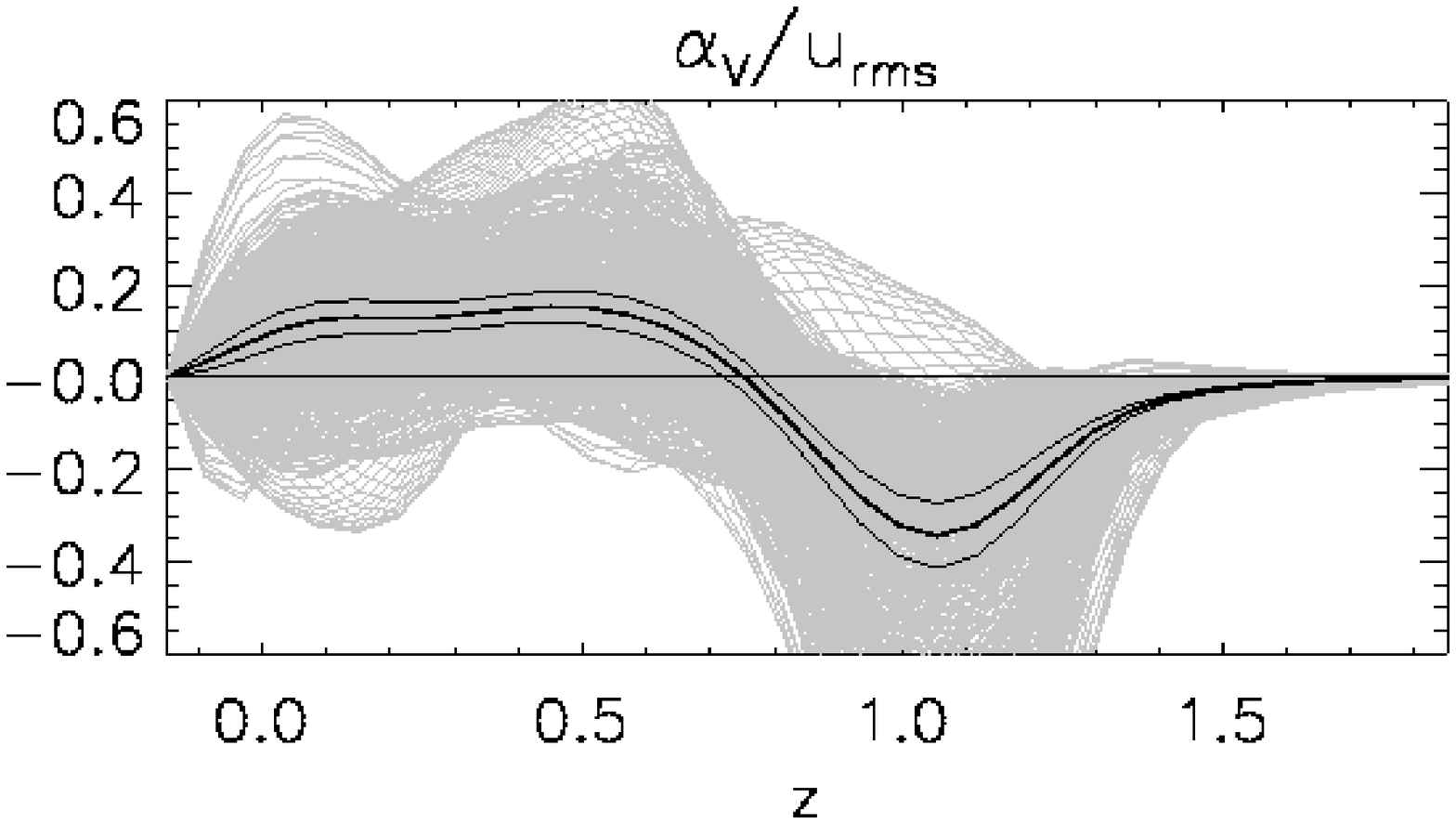,width=6cm }
\psfig{file=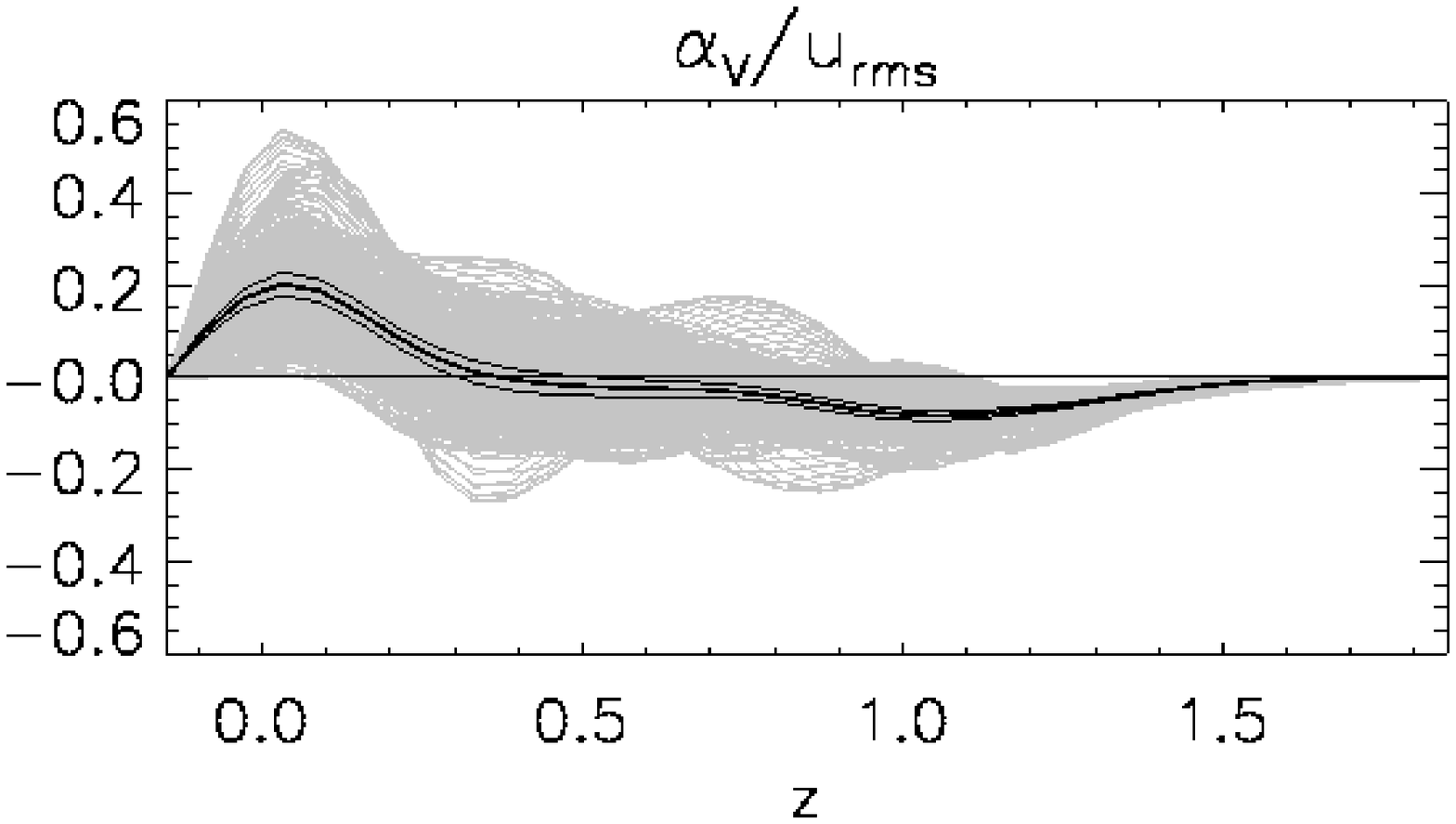,width=6cm }
\psfig{file=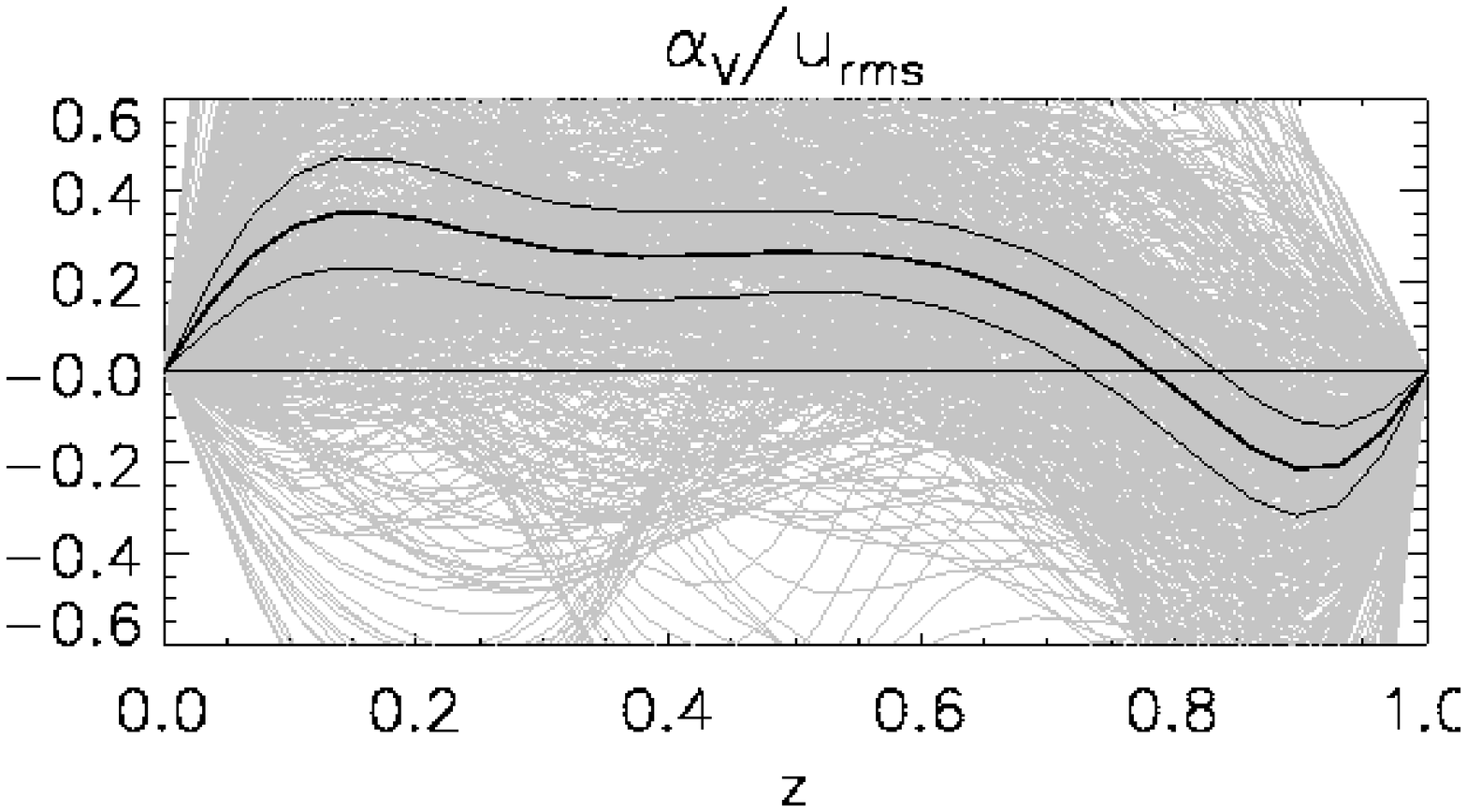,width=6cm }
}
\centerline{
\psfig{file=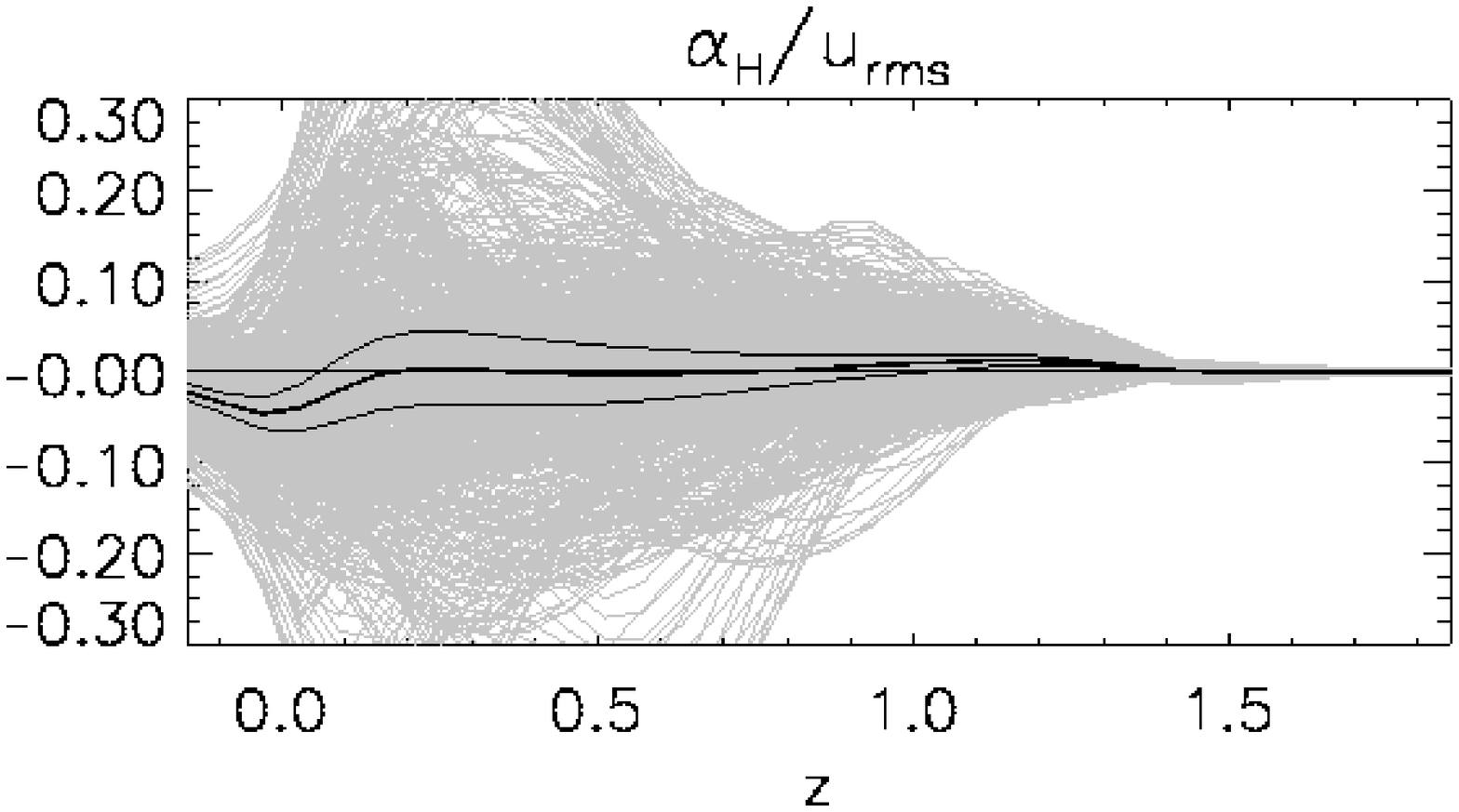,width=6cm }
\psfig{file=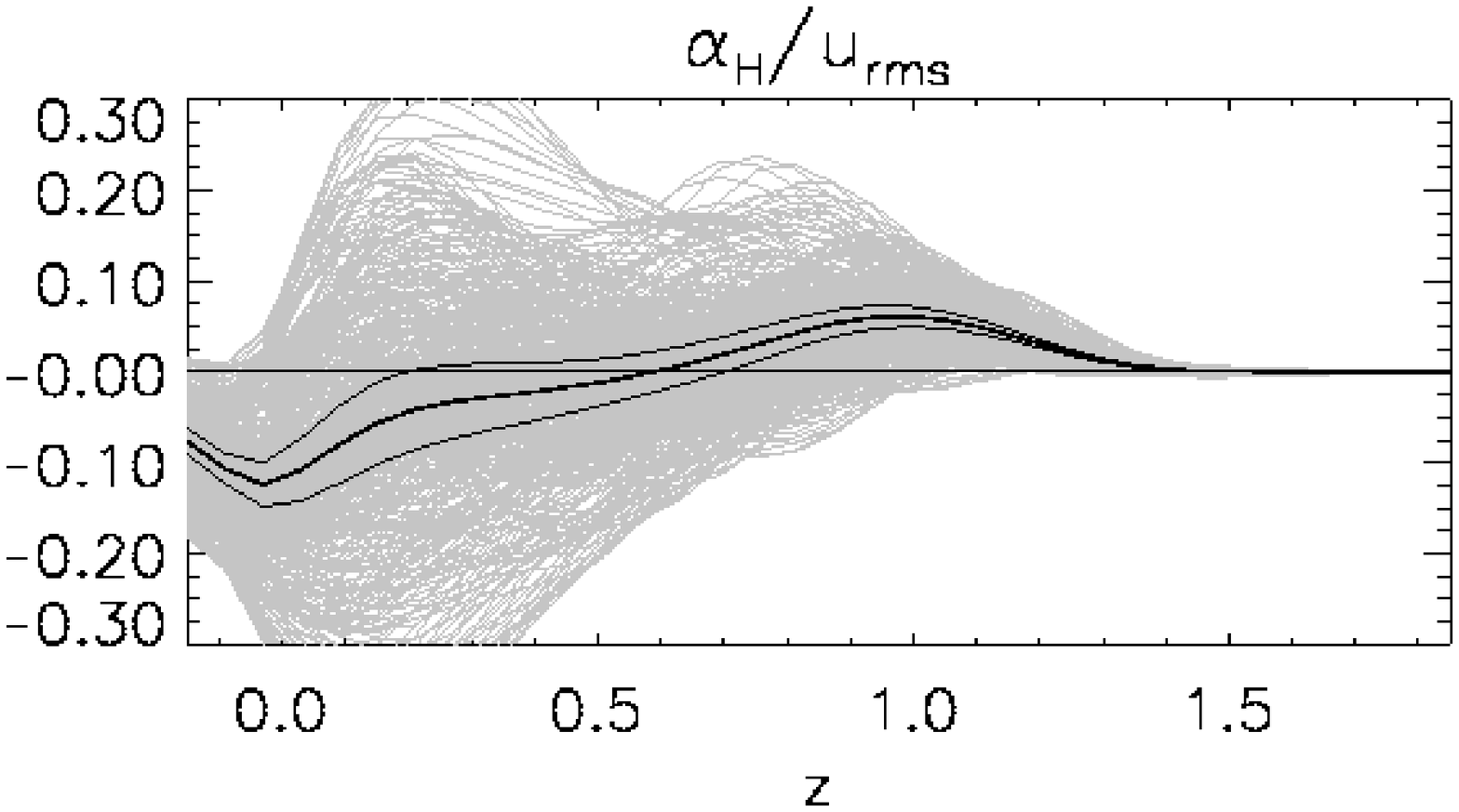,width=6cm }
\psfig{file=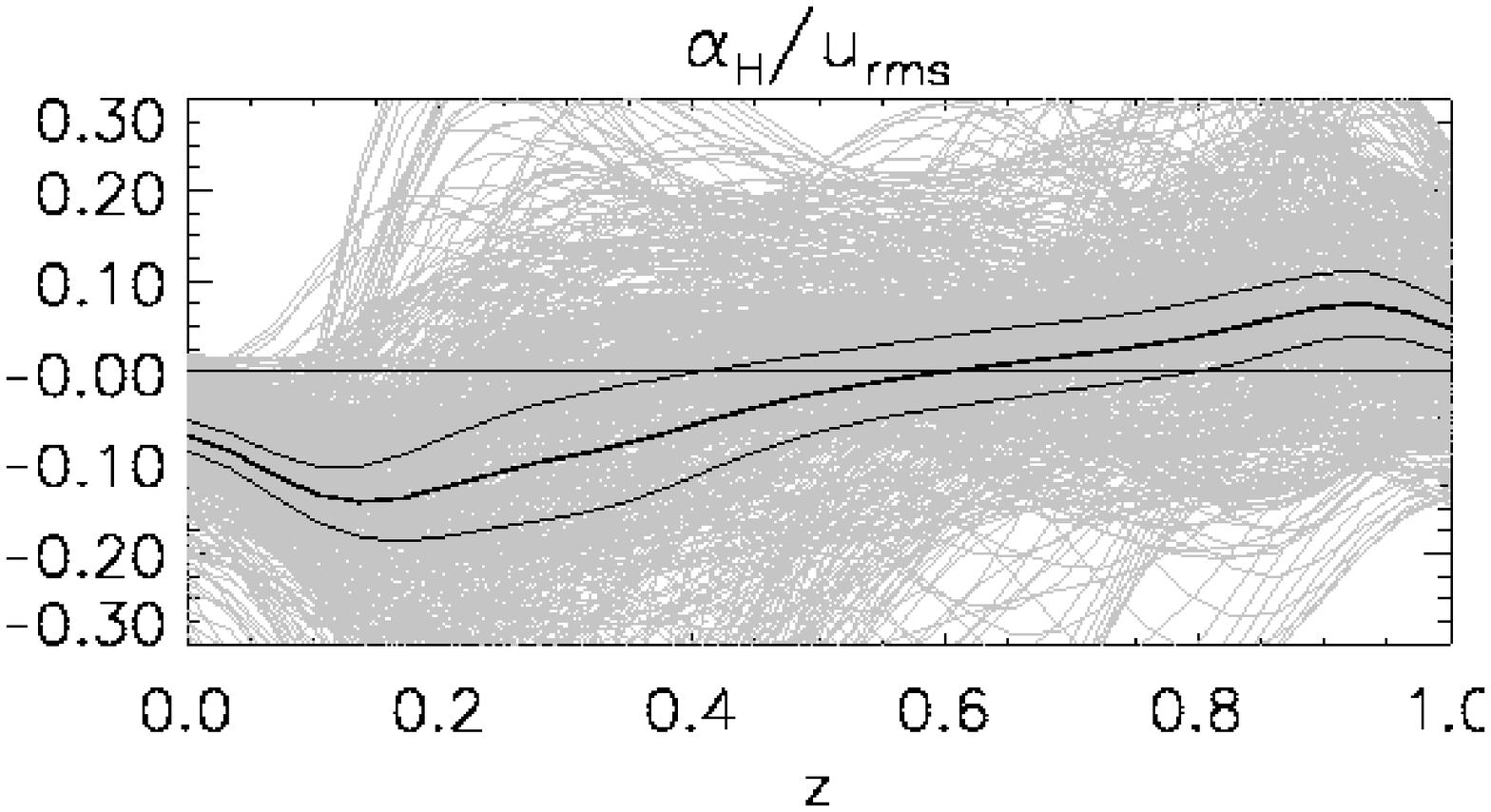,width=6cm }
}
\centerline{
\psfig{file=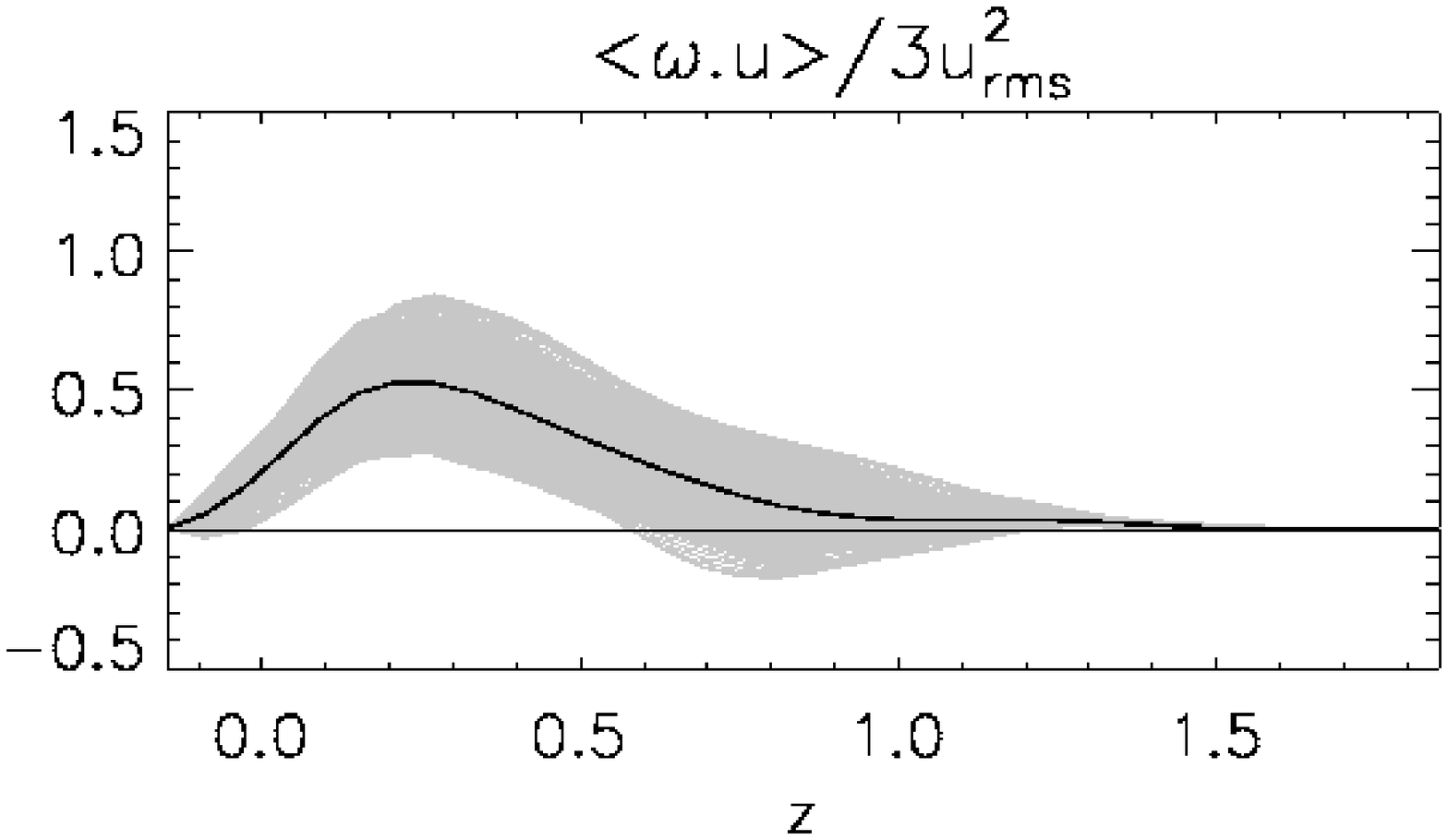,width=6cm }
\psfig{file=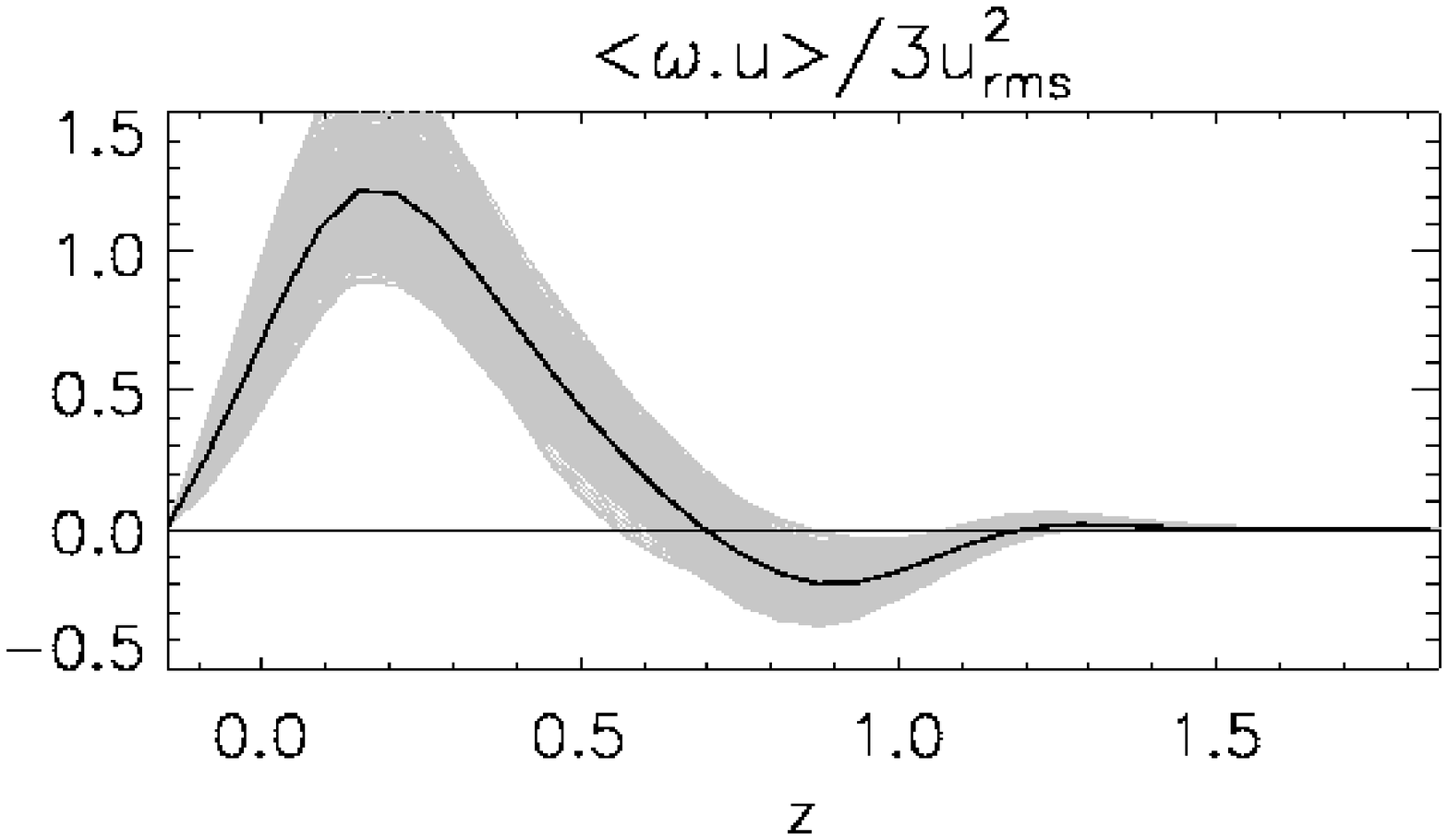,width=6cm }
\psfig{file=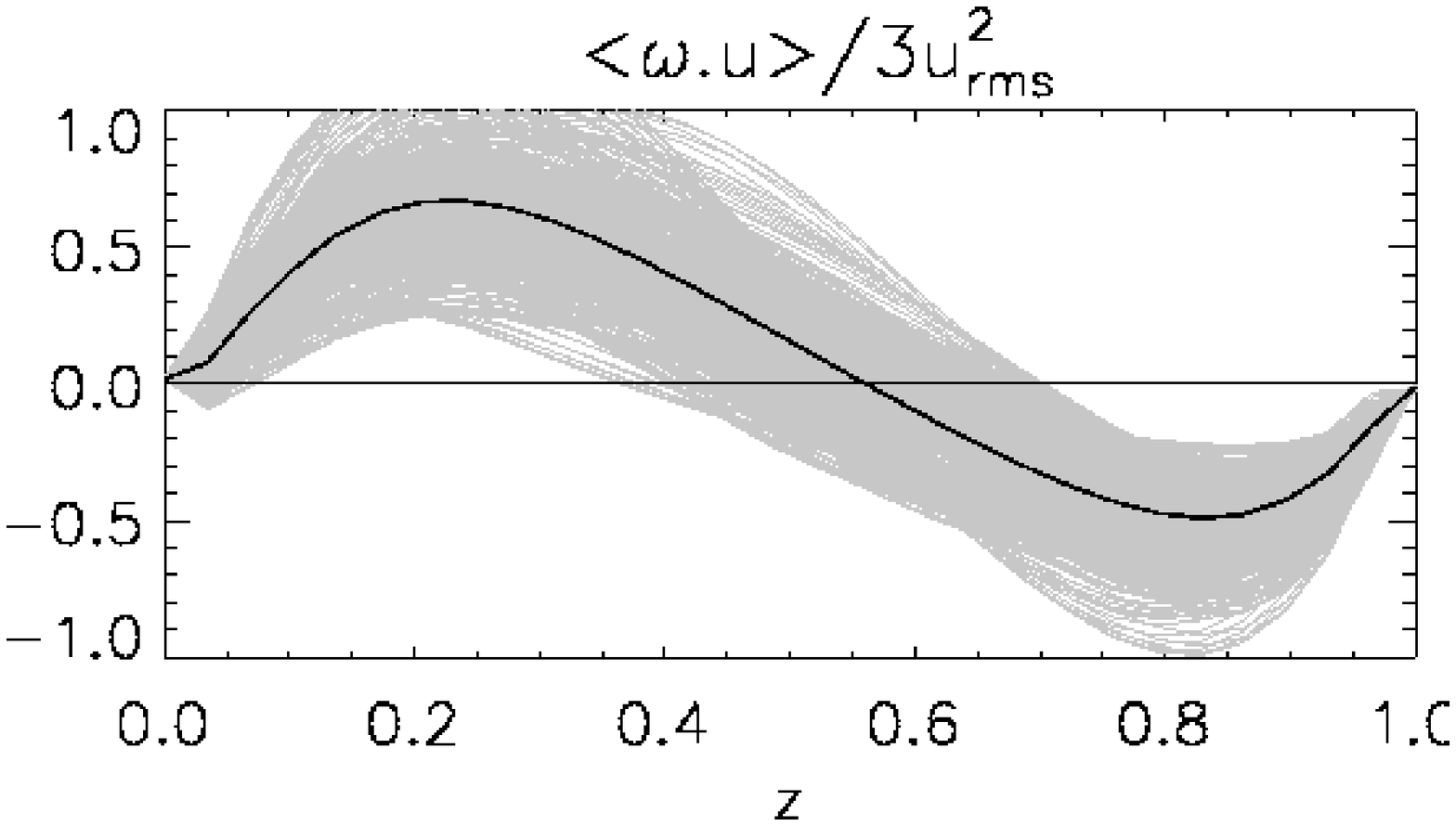,width=6cm }
}
\centerline{
\psfig{file=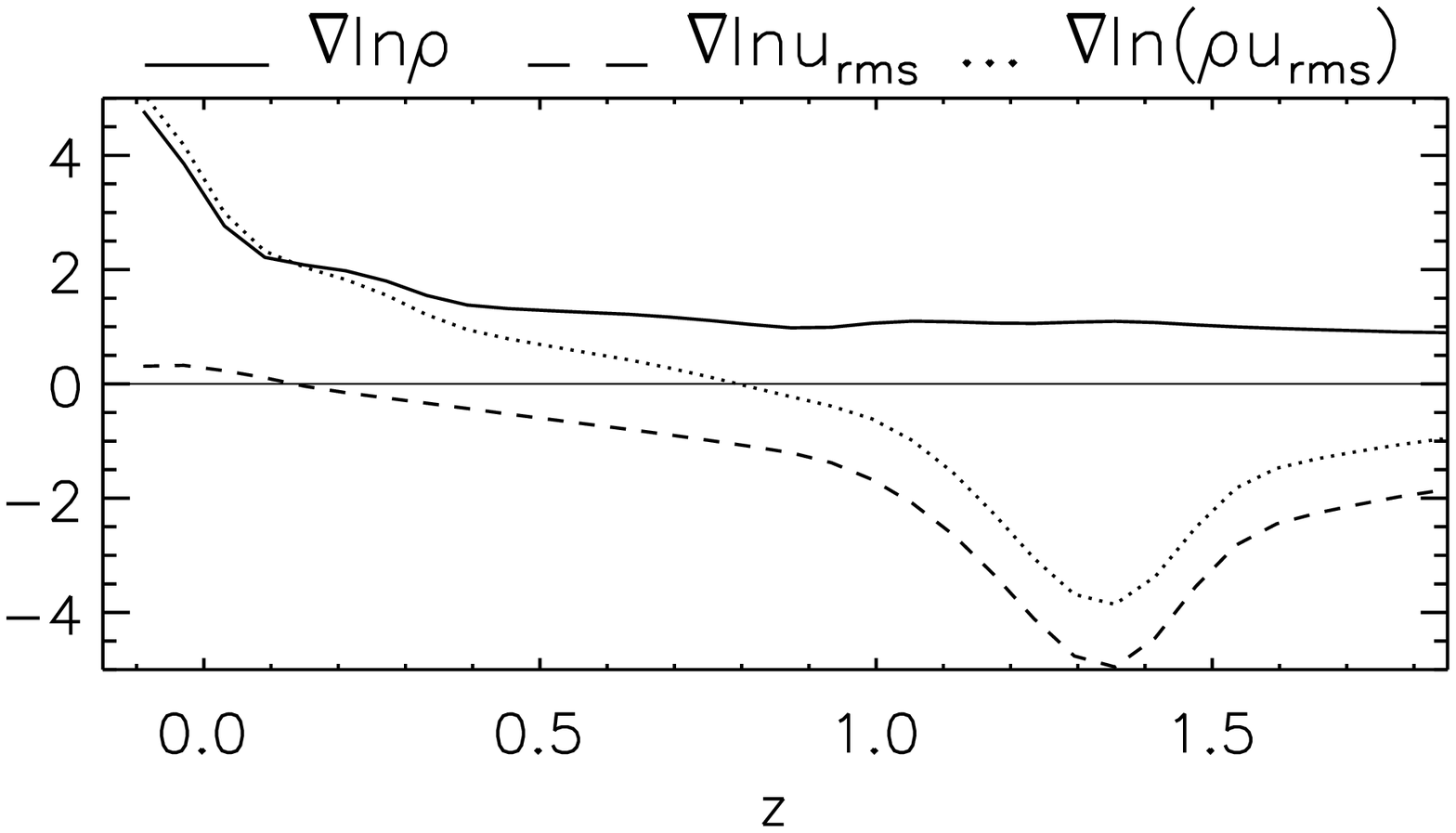,width=6cm }
\psfig{file=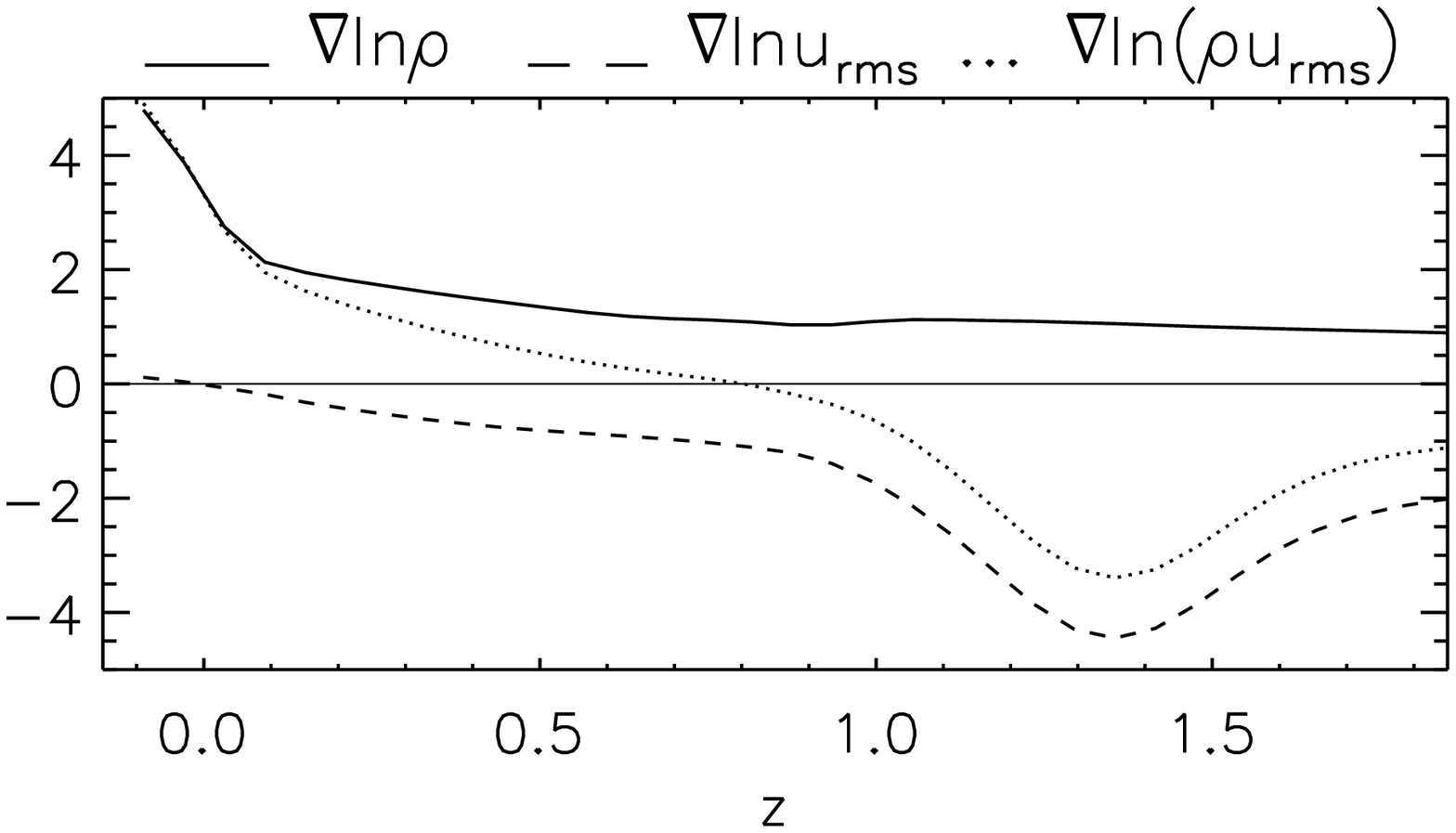,width=6cm }
\psfig{file=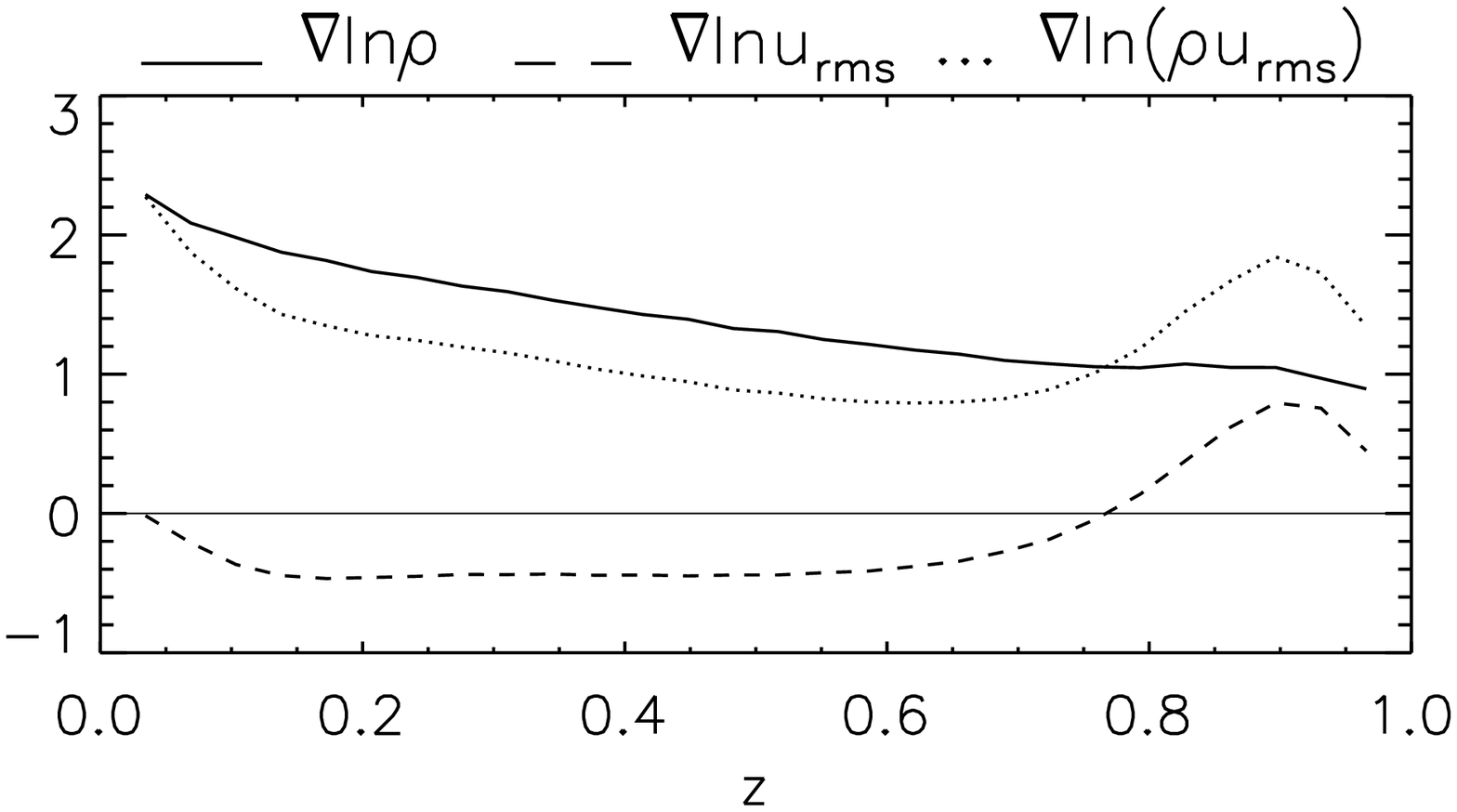,width=6cm }
}
\centerline{
\psfig{file=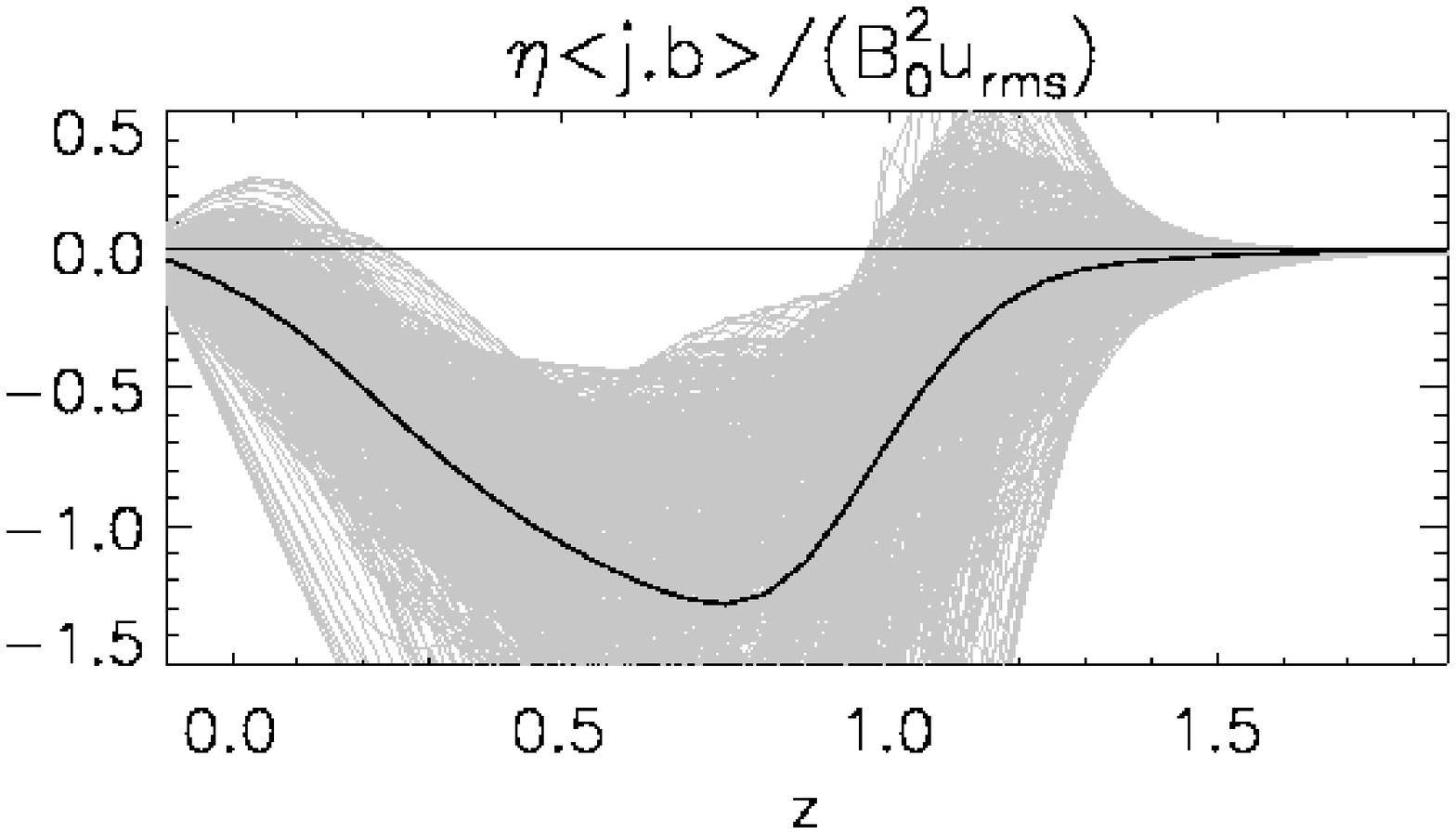,width=6cm }
\psfig{file=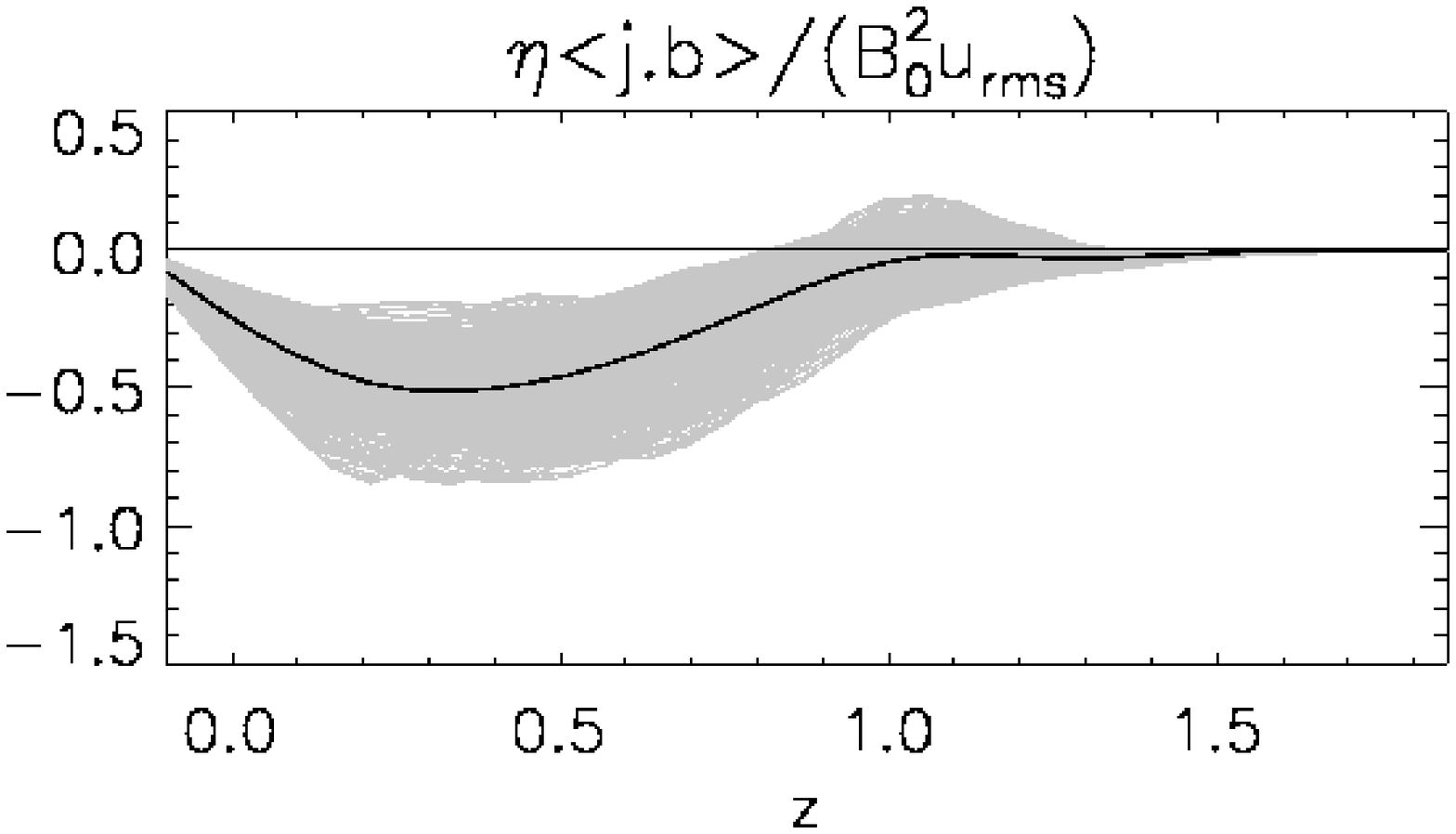,width=6cm }
\psfig{file=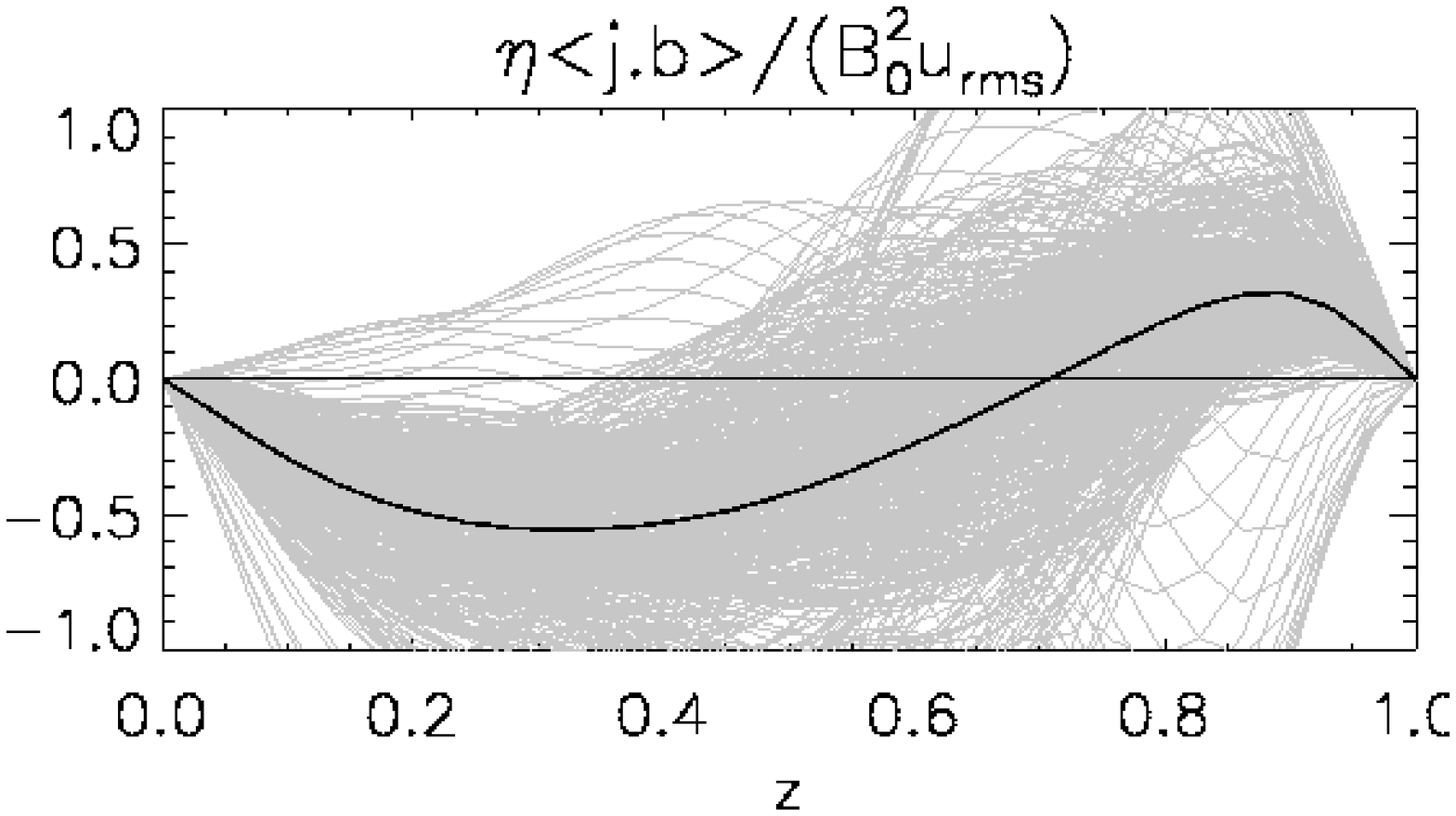,width=6cm }
}
\caption{{\em Row 1}\/: vertical $\alpha$ effect. {\em Row 2}\/: 
         horizontal $\alpha$ effect. {\em Row 3}\/: kinetic helicity. 
         {\em Row 4}\/: gradients of $\ln\rho$ and $\ln u_{\rm rms}$, and
         their sum (rows 3 and 4 are for a vertical imposed field, but
         since the field is weak the results in these rows would be the
         same for a horizontal field). 
         {\em Row 5}\/: current helicity for a vertical imposed
         magnetic field. {\em Left column}\/: run 6 ($\mb{Ta}=2000$).
         {\em Middle column}\/: run 7 ($\mb{Ta}=10\,000$). 
         {\em Right column}\/: run 10 ($\mb{Ta}=10\,000$; no stable layers).
         In all cases the unstable region is between $z=0$ and $z=1$.
         All calculations are for the solar south pole.
         Error margins are given for rows 1 and 2 as thin lines 
         enclosing the mean; those in rows 3--5 would be comparable or
         narrower, but are omitted in order not to overload the 
         drawings.}                                     \label{pkinhel}
\end{figure*} 

\subsection{Statistics}
It is well known from other numerical simulations that $\alpha$ is an 
extremely noisy quantity, so that one should do as much averaging as 
possible in order to optimize the statistics. 

For this paper we have chosen two averaging procedures. The first,
which is considered in the present section, consists of an average over
the horizontal coordinates and over time (Fig.~\ref{fig2}, upper
panel). The time average is initiated from the moment when the
magnetoconvection has attained a statistically stationary state, i.e.
when the kinetic and magnetic energy densities are more or less constant.
The second is an additional average  over two suitably chosen depth ranges
(Fig.~\ref{fig2}, lower panel): since the $\alpha$ effect depends on
depth, and is expected to change sign near the bottom of the unstable layer,
the averaging  over depth is performed separately for the unstably and the 
stably stratified layers. This second procedure will be used mainly
in later sections.

The statistical deviation of $\alpha$ from its mean value generally
is large, often larger than the difference between that mean and zero,
as illustrated in Fig. \ref{pkinhel}. Therefore, extended simulations
are necessary in order to obtain a significant result. In the present
paper the error estimate is based on the assumption that a single
simulation can be divided into a number of time intervals containing
independent realizations of the small-scale flow and magnetic field.
The length $\tau$ of those time intervals should then exceed the
correlation time, which is $\approx 20$ (in dimensionless units), as
determined from the width of the autocorrelation function. To be on
the safe side, we smoothed curves of $\alpha_{\rm V\!,unst}(t)$
and $\alpha_{\rm V\!,stab}(t)$ for several runs by performing a box
average, and then calculated the autocorrelation function. This
procedure yields a typical coherence time $\tau=50$, which is several
times longer than the value derived from the unsmoothed curves.

Our procedure of estimating errors is more primitive than that applied
earlier to results of numerical simulations (Pulkkinen et al. 1993).
Moreover, the errors of the mean values determined in this manner can
only be crude estimates because firstly we cannot be sure that we deal
with Gaussian statistics (although inspection of the examples of Fig. 
\ref{pkinhel} suggests that $\alpha$ varies in a nearly symmetric manner
around its mean), and secondly the number $T/\tau$ of time intervals
is not very large. Nevertheless we think that the procedure indicates
whether or not the obtained mean values are significantly different
form zero. Test simulations with identical parameters but different
initializations confirm this indication. Of course, the resulting
error bars do not imply anything about the physical assumptions made.
In particular, we cannot expect that models based on different
assumptions, such as ours and that of R\"udiger \& Kitchatinov
(\cite{ruediger93}), will yield results that lie within error bars
obtained from statistics (these bars shrink to zero for
$T \rightarrow \infty$). In contrast, we must see whether we can
find qualitatively similar or dissimilar results from different
models, especially as none of the models yet meets the real Sun.

\subsection{Relations between $\alpha$ and turbulence properties}
The existence of the $\alpha$ effect is attributed to the helical 
nature of convective flows in a rotating medium. Figure \ref{pkinhel}
shows $\alpha_{\rm V}$, $\alpha_{\rm H}$ and several quantities that,
as is shown below, have a relation to the $\alpha$ effect. 
Three different runs are selected in order to demonstrate 
the effects of rotation and boundary conditions. 

The $\alpha$ coefficients are depth-dependent, 
and undergo a sign change within the unstable layer.
For $\mb{Co}\la 4$, the vertical alpha coefficient exceeds the 
horizontal coefficient, and has the opposite sign. If rotation is 
stronger, the vertical $\alpha$ effect is strongly reduced.

For isotropic turbulence, the first-order smoothing approximation 
(FOSA) yields that $\alpha$ can be described by a single scalar
\begin{equation}
\alpha\approx-\frac{1}{3}\tau\lb\vec{\omega}\cdd\vec{u}\rb\,, \label{alp1}
\end{equation}
where $\tau$ is the correlation time of the convection, 
and $\lb\vec{\omega}\cd\vec{u}\rb$ 
is the kinetic helicity (Krause \& R\"adler \cite{krause80}). 
An alternative representation, valid under the assumption of FOSA, relates $\alpha$
to the {\em current helicity} (Keinigs \cite{keinigs83}; 
R\"adler \& Seehafer \cite{raedler90};
note the extra minus sign in Keinigs' definition of $\alpha$):
\begin{equation}
\alpha\approx -\eta\,\lb\vec{j}\cdd\vec{b}\rb/|\lb\vec{B}\rb|^2\,.\label{alp2}
\end{equation}
It should be noted that (\ref{alp2}) does not reflect the nonlinear feedback of the Lorentz force
on the flow, but is a purely linear result.
If one allows for mild anisotropy in the radial direction, alpha assumes the form of a tensor, and a 
detailed calculation (Steenbeck \& Krause \cite{steenbeck69}) reveals that, for slow rotation 
($\mb{Co}\ll 1$), the diagonal term is given by 
\begin{equation}
\alpha\approx -\frac{16}{15}\tau^2u_{\rm rms}^2
              \vec{\Omega}\cdd\na\ln\, (\rho u_{\rms})\,. \label{alp3}
\end{equation}
While conditions under which expressions (\ref{alp1})--(\ref{alp3}) hold are fulfilled neither in the simulations 
nor in the Sun (mainly because of the requirements for a short correlation time and only mild 
anisotropies in the turbulence), it is still instructive to compare them with the numerical results.

The kinetic helicity is positive at the top of the convective layer (Fig. \ref{pkinhel}, row 3), as 
is expected for the southern hemisphere based on the direction of the Coriolis force for contracting 
sinking parcels, or expanding rising parcels. Brummell et al. (\cite{brummell98}) find kinetic 
helicity with the same sign for a box on the {\em northern} 
hemisphere, but their coordinate system turns out to have a left-handed orientation. Near the center 
of the convective layer, the sign of the kinetic helicity reverses if rotation is sufficiently strong, 
because below a certain point sinking parcels expand laterally while 
approaching the level $z=1$ (i.e., the stably stratified overshoot layer or
the impenetrable lower boundary if an overshoot layer is absent), 
while rising parcels contract while moving away from it. For $\mb{Co}\la 2$,
no reverse-helicity region seems to exist. Nevertheless, in these cases
alpha still changes its sign near the bottom of the convective layer. For
comparison we have calculated a case (run 10) with the same Taylor number as
in run 7, but where the stably stratified layers are replaced by impenetrable
boundaries at $z=0$ and $z=1$. In this case the kinetic helicity near the
bottom of the unstable layer is more strongly negative, and has a magnitude 
comparable to that in the upper part of the unstable layer (Fig. \ref{pkinhel},
row 3, col. 3). This can be attributed to the fact that an impenetrable 
boundary forces a sinking parcel to diverge more strongly than does a 
stably stratified layer. Thus, except for the stable region in some cases, 
the sign of $\alpha$ as predicted by Eq. (\ref{alp1}) roughly agrees with
that of $\alpha_{\rm H}$, while it is opposite to that of $\alpha_{\rm V}$. 
The deviating sign of $\alpha_{\rm V}$ was explained by Brandenburg et al.
(\cite{brandenburg90}). We note that the same unconventional sign of 
$\alpha_{\rm V}$ was also found by Ferri\`ere (\cite{ferriere92}), and 
R\"udiger \& Kitchatinov (\cite{ruediger93}), although not for the same
values of the Coriolis parameter as in our simulations.

The gradient of $\ln\,(\rho u_{\rm rms})$ is positive at the top of the
convective layer and changes sign at a point near the bottom of the 
unstable layer in the runs with an overshoot layer (Fig. \ref{pkinhel}, 
row 4, cols. 1,2). Hence the sign of $\alpha$ as predicted by (\ref{alp3})
agrees with that of $\alpha_{\rm H}$ in the simulations, also if rotation
is weak. In this respect, $\na\ln\,(\rho u_{\rm rms})$ agrees qualitatively
more closely than does the kinetic or the current helicity. In the upper 
region, $\na\ln\,(\rho u_{\rm rms})$ is positive because the density 
gradient dominates, while near the bottom of the unstable domain the 
influence of the convective stability is felt, resulting in a reduction 
of the turbulent velocity sufficient to produce a negative sign.
With regard to $\alpha_{\rm V}$, the sign and the dependence on Co
differ strongly from (\ref{alp3}).
Also, in the run without the stably stratified layer, the 
gradient of the density dominates throughout the box, and no sign change of
$\na\ln\,(\rho u_{\rm rms})$ is observed (Fig. \ref{pkinhel}, row 4, col. 3). 
Both components of $\alpha$ do exhibit a sign change, though. This 
inconsistency may be a result of the impenetrable boundary condition at
$z=1$, which causes a transfer of kinetic energy from the vertical 
to the horizontal components, an effect which is not accounted for by 
Eq. (\ref{alp3}). A detailed comparison of the rotational dependence of
$\alpha_{\rm V}$ and $\alpha_{\rm H}$ with Eq. (\ref{alp3}), and with the 
analytical results of R\"udiger \& Kitchatinov (\cite{ruediger93}), will
be presented in Sect. 4.2.

\begin{figure}[b]
\centerline{\psfig{file=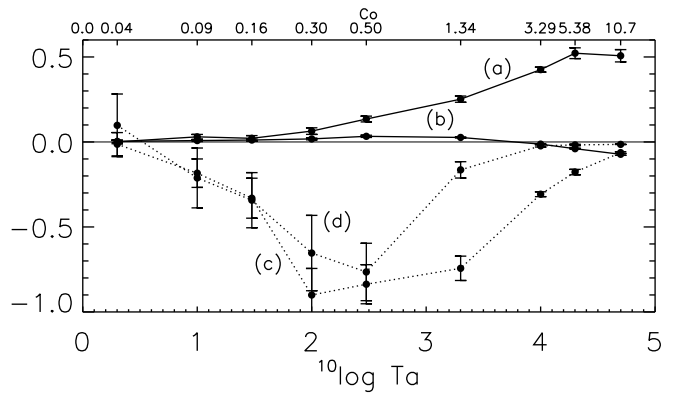,width=8.9cm}}
\caption{{\em Solid}\/: Kinetic helicity,
         $\lb\vec{u}\cdd\vec{\omega}\rb d/3u_{\rm rms}^2$; 
         {\em dotted}\/: current helicity, 
         $\eta\,\lb\vec{j}\cdd\vec{b}\rb/B_0^2 u_{\rm rms}$,
         for a vertical imposed magnetic field at the solar south pole. 
         ({\em a,c}) average over $-0.15\leq z\leq 1$;
         ({\em b,d}) average over $1\leq z\leq 1.5$.}     \label{ptahel}
\end{figure}

The current helicity is negative in the upper part of the unstable layer
for a vertical imposed magnetic field (Fig. \ref{pkinhel}, row 5), which
confirms earlier results of Brandenburg et al. (\cite{brandenburg90}). A
reverse-polarity layer is not observed, except in run 10 with the 
impenetrable boundary at $z=1$. Thus, the sign of $\alpha$
in the unstable layer as predicted
by Eq. (\ref{alp2}) is the same as that of $\alpha_{\rm V}$. The negative
sign of $\lb\vec{j}\cd\vec{b}\rb$, which corresponds to a positive sign
in the bulk of the convection zone on the northern hemisphere, is at odds
with forced isotropic turbulence calculations by Brandenburg 
(\cite{brandenburg00}) as well as with observations of the solar surface
(Seehafer 1990, Low \cite{low96}) and mean-field calculations (R\"adler \&
Seehafer \cite{raedler90}, R\"udiger et al. \cite{ruediger01}). For a
horizontal imposed magnetic field, the current helicity is about one order
of magnitude smaller than in the case of a vertical imposed field. It is 
highly fluctuating, and its mean value can have either sign (not shown). 
The smaller amplitude is consistent with the smaller amplitude of the
magnetic fluctuations (Fig. \ref{purms}).

\section{The $\alpha$ Effect as a Function of $\Omega$}

In this section we investigate the dependence of the $\alpha$ effect
on the angular velocity $\Omega$. The relevant dimensionless parameters
are the Taylor number and the Coriolis number, as defined by (22) and
(23). Averages are calculated over time, and separately over the unstable
region and the stable region, as explained above.

\subsection{Kinetic Helicity and Current Helicity}

\begin{figure}[thb]
\centerline{\psfig{file=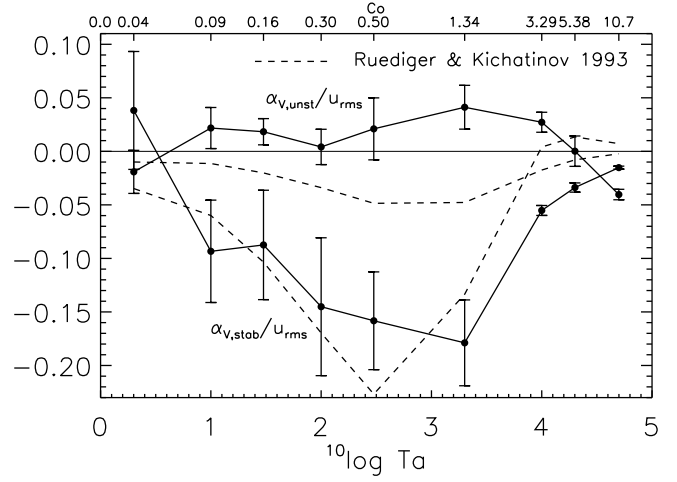,width=8.7cm}}
\centerline{\psfig{file=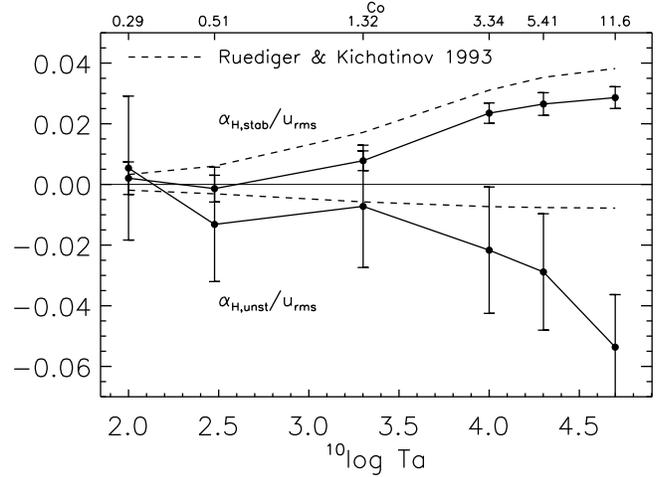,width=8.7cm}}
\caption{$\alpha$-coefficients, normalized by $u_{\rms}$, as functions 
         of Ta. Here $u_{\rms}$ is an average over time and over the 
         unstable layer. {\em Top}\/: vertical $\alpha$ effect. 
         {\em Bottom}\/: horizontal $\alpha$ effect. The errors are 
         standard deviations divided by the square root of the number 
         of turnover times covered by the simulation. All simulations
         are for the solar south pole. The dashed curves 
         are analytical results based on  R\"udiger \& Kichatinov 
         (\protect{\cite{ruediger93}}). In each figure, a suitable uniform
         scaling factor was applied to the analytical curves.}   \label{pta}
\end{figure}

\begin{figure}
\centerline{\psfig{file=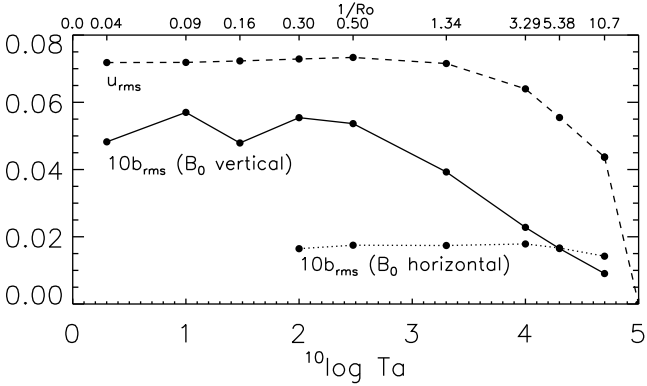,width=8.8cm}}
\caption{Strength of the velocity and magnetic field perturbations as a 
         function of Ta, in units of $\sqrt{gd}$, and $\sqrt{\mu_0\rho_0 gd}$,
         respectively. The vertical and horizontal orientation of the imposed
         magnetic field are used for $\alpha_{\rm V}$ and
         $\alpha_{\rm H}$, respectively.}                    \label{purms}
\end{figure}

Figure \ref{ptahel} shows the rotational dependence of the kinetic
helicity and the current helicity (for the vertical imposed magnetic 
field). The kinetic helicity and the current helicity
have been multiplied with $d/3u_{\rm rms}^2$ and $\eta/B_0^2 u_{\rm rms}$,
respectively, in order to allow a comparison with $\alpha/u_{\rm rms}$,
as suggested by Eqs. (\ref{alp1}) and (\ref{alp2}) if one sets 
$\tau\approx d/u_{\rm rms}$. The kinetic helicity in the unstable layer
increases with the rotation rate up to $\mb{Ta}\approx 10^4$; its value
in the stable layer is much smaller. Hence the general trend and 
the correct sign of $\alpha_{\rm H}$  (Fig. \ref{pta}, bottom) are
recovered by (\ref{alp1}) in the unstable layer, but not in the stable layer. 
The current helicity for the vertical imposed magnetic field increases 
with rotation up to $\mb{Ta}\approx 100-300$, while it decreases for
stronger rotation, as do the magnetic fluctuations (Fig. \ref{purms}).
It seems that $\alpha_{\rm V}$ in the unstable layer roughly traces the 
current helicity (Figs. \ref{ptahel} and \ref{pta}). 
Note however that both the normalized kinetic helicity,
$\lb\vec{\omega}\cdd\vec{u}\rb/\omega_{\rm rms}u_{\rm rms}$, and the 
normalized current helicity, $\lb\vec{j}\cdd\vec{b}\rb/j_{\rm rms}b_{\rm rms}$,
increase monotonically with $\mb{Ta}$ (not shown). 
This reflects an increasing degree of alignment of vorticity with velocity,
and of current with magnetic field, respectively. 

The main conclusion to be drawn with regard to the validity of
(\ref{alp1})--(\ref{alp3}) is that qualitative agreement exists only
between $\alpha_{\rm H}$, (\ref{alp1}), and (\ref{alp3}) on the one hand,
and between $\alpha_{\rm V}$ and (\ref{alp2}) in the unstable layer on
the other hand. 
For $\mb{Co}\ga 5$, such qualitative similarities are no longer evident,
but in the following sub-section we shall compare
the behavior of $\alpha_{\rm V}$ and $\alpha_{\rm H}$ with mean-field
results that were derived for the case of arbitrary rates of rotation.

\subsection{Rotational quenching of $\alpha$}
Figure \ref{pta} shows $\alpha_{\rm V}$ and $\alpha_{\rm H}$, normalized
by $u_{\rms}$, as a function of Ta (or Co). For zero rotation the
$\alpha$ effect vanishes. For weak rotation $\alpha_{\rm V}/u_{\rm rms}$
increases with increasing rotation rate. It reaches a maximum near 
$\mb{Co}\approx 1.3$ ($\mb{Ta}\approx 2\cd 10^3$); for stronger rotation
it is quenched. The decrease of $\alpha_{\rm V}/u_{\rm rms}$ occurs in
spite of the monotonic increase of the kinetic helicity, and in spite of
the decrease of $u_{\rms}$ with increasing rotation rate (Fig. \ref{purms}).
The horizontal $\alpha$ effect sets in at a higher rotation rate.
Its magnitude increases monotonically with the rotation rate, 
and exceeds that of the vertical
$\alpha$ effect if $\mb{Co}\ga 8$. For $\mb{Ta}=10^5$, Coriolis forces
are so strong that they stifle the convection (see below), and hence also
alpha. Two effects that are not shown in detail appear to be responsible
for the reduction of $\alpha_{\rm V}$: an increasing number of sign changes
of $\alpha_{\rm V}(z)$, resulting in cancelations when the volume average
is calculated, and a decreasing amplitude of the magnetic fluctuations 
(Fig. \ref{purms}). Both effects are absent in the case of $\alpha_{\rm H}$.

\begin{figure}
\centerline{\psfig{file=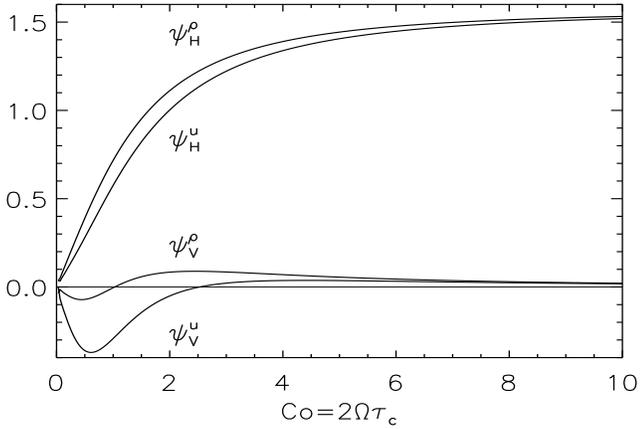,width=8.8cm,height=6cm}}
\caption{Functions $\psi_{\rm V,H}^{\rho,u}$, which describe the Coriolis
         number dependence of contributions to $\alpha_{\rm V}$ and 
         $\alpha_{\rm H}$ due to density and turbulence stratification, 
         respectively.} \label{prk}
\end{figure}

Clearly, the alpha effect is highly anisotropic, and its dependence on 
rotation is more complicated than suggested by Eqs. 
(\ref{alp1}) and (\ref{alp3}). R\"udiger \& Kichatinov (\cite{ruediger93}) 
present analytical expressions for the rotational dependence of 
$\alpha_{\rm V}$ and $\alpha_{\rm H}$, based on the first-order smoothing
approximation (FOSA), but for arbitrary rotation rates.
They employ a linearized equation of motion for the fluctuating quantities
which includes a random driving force with prescribed spectral properties
to model turbulence, and which takes into account a stratification of the 
density and of the turbulent velocity. Due to the stratification as well 
as rotation, the electromotive force becomes anisotropic. In the simplest
case of a driving force with zero frequency and containing a single 
wavelength (the case designated mixing-length approximation by the authors),
and for a weak magnetic field, a manageable expression can be derived. 
For the present geometry, i.e., a local Cartesian grid situated at the 
south pole, one obtains
\begin{eqnarray}
\alpha_{\rm V}&=&-\tau u_{\rms}^2 \Big[ \psi_{\rm V}^{\rho}\na\ln\rho +
  \psi_{\rm V}^u\na\ln u_{\rms}\Big] \label{alphavrk}\\
\alpha_{\rm H}&=&-\tau u_{\rms}^2\Big[ \psi_{\rm H}^{\rho}\na\ln\rho +
  \psi_{\rm H}^u\na\ln u_{\rms} \Big]. \label{alphahrk}
\end{eqnarray}
These expressions correspond to those of R\"udiger \& Kichatinov 
(\cite{ruediger93}) if one sets $\alpha_{\rm V}=\alpha_{zz}$,  
$\alpha_{\rm H}=\alpha_{xx}$, $\psi_{\rm H}^{\rho}=\mb{Co}\,\Psi^{\rho}$,
and $\psi_{\rm H}^u=\mb{Co}\,\Psi^u$ (cf. their Eq. 3.11). 
The functions $\psi_{\rm V}^{\rho,u}$ are given by
\begin{eqnarray}
\lefteqn{\psi_{\rm V}^{\rho} =\frac{1}{x^3}\Big[\frac{1}{2}x^2-\frac{23}{2}
               -\frac{(x^2-1)^2}{2(x^2+1)}}\nonumber\\
               &&\hspace*{3cm}+2\frac{x^2+6}{x}\arctan x \Big]\\
\lefteqn{\psi_{\rm V}^u=\frac{1}{x^3}\Big[ x^2-17-\frac{(x^2-1)^2}
       {x^2+1}}\nonumber\\
               &&\hspace*{3cm}+2\frac{x^2+9}{x}\arctan x\Big],
\end{eqnarray}
where $x=\mb{Co}$. Several features relevant for our discussion should 
be noted. As is shown in Fig. \ref{prk}, the functions 
$\psi_{\rm H}^{\rho,u}$ are almost equal and, for $\mb{Co}\la 3$, they 
are linear in $\mb{Co}$, so that Eq. (\ref{alphahrk}) reduces to 
Eq. (\ref{alp3}), apart from a factor of order unity.
Secondly, $\psi_{\rm V}^{\rho,u}$, and therefore $\alpha_{\rm V}$, are 
subject to rotational quenching. This unexpected behavior, which is not 
predicted by Eqs. (\ref{alp1})--(\ref{alp3}), is roughly confirmed by the
simulations, as is shown by Fig. \ref{pta}. In order to produce the 
dashed curves, representative values, estimated from the simulations, were
inserted for the gradients $\na\ln\rho$ and $\na\ln u_{\rms}$, namely $1.5$
and $-1.0$ in the unstable layer, and $1.0$ and $-3.5$ in the stable layer
(Fig. \ref{pkinhel}, row 4).
For a given Taylor number, the same value of $\mb{Co}$ was used for both
the unstable layer and  the stable layer, i.e. the turnover time, 
$\tau\approx\ell/u_{\rms}$, was assumed to be the same. This is motivated
by the fact that both the effective thickness, $\ell$, and $u_{\rms}$ for the 
stable layer are smaller by a factor 3--4 compared to their values in the
unstable layer. Secondly, the temporal variations of the flow are very
similar in both layers (Fig. \ref{fig2}, bottom). After dividing Eqs.
(\ref{alphavrk}) and (\ref{alphahrk}) by $u_{\rms}$, a factor $\tau u_{\rms}$
remains in front which can be assumed to be independent of $\mb{Co}$.
However, it was found that setting this factor to $\tau u_{\rms}=\ell$
overestimates the amplitude of $\alpha$. For the dashed curves shown in
Fig. \ref{pta}, it was replaced by a suitable uniform scaling factor, 
namely $0.19$ for $\alpha_{\rm V}$ and $0.01$ for $\alpha_{\rm H}$. 
Thus, our simulations produce much smaller $\alpha$ coefficients than
predicted by (\ref{alphavrk}) and (\ref{alphahrk}), especially 
$\alpha_{\rm H}$ which is smaller by two orders of magnitude.
On the other hand, the main point is the agreement with regard to the
rotational dependence. Except for the sign of $\alpha_{\rm V}$ in the 
unstable layer, the rotational quenching is reproduced, including the
position of the maxima. For weak 
rotation, $\psi_{\rm V}^u\approx 4\psi_{\rm V}^{\rho}$, so that the 
contribution of $\na\ln u_{\rms}$ dominates, while for $\mb{Co}\ga 2$ the
density gradient dominates; in both cases the mean-field formula yields
the wrong negative sign. Clearly, the sign of $\alpha_{\rm V}$ as
predicted by Eq. (\ref{alphavrk}) can have both values in principle, and
depends in a delicate way on the relative importance of the two gradients;
even a slight change of the $\psi$ coefficients can generate a sign
chance. With regard to $\alpha_{\rm H}$, we confirm the saturation
predicted by Eq. (\ref{alphahrk}) only in the stably stratified layer,
but not (yet) in the unstable layer (Fig.~\ref{pta}, bottom).
Apart from this, the general trend as well as the correct signs are 
reproduced by (\ref{alphahrk}) in both layers. Due to the shape of the
functions $\psi_{\rm H}^{\rho,u}$, this feature should be rather robust.
 
\begin{table*}[htb]
\caption{Eddy sizes and preferred length scales for the onset of the convective
         instability between two free boundaries as given by Chandrasekhar
         (\protect{\cite{chandra61}}). The eddy sizes were determined visually;
         their random error is estimated to be about $25\%$.}\label{tablesize}
\begin{center}
\begin{tabular}{ccccccccccc} \hline\noalign{\smallskip}
run             & 1  & 2  & 3  & 4  & 5  & 6   & 7     & 8          &  9       \\
\noalign{\smallskip}\hline\noalign{\smallskip}
$\mb{Ta}$       & 2  & 10 & 30 & 100& 300& 2000& $10^4$& $2\cd 10^4$& $5\cd 10^4$\\ 
eddy size       & 2.5& 2.5& 2.5& 2.5& 2.0& 1.5 & 1.3   & 1.0        & 0.8 \\
$2\pi/k_{\rm c}$& 2.8& 2.8& 2.7& 2.4& 2.1& 1.5 & 1.1   & 1.0        & 0.82\\
\noalign{\smallskip}\hline
\end{tabular}
\end{center}
\end{table*}

\begin{figure}
\centerline{\psfig{file=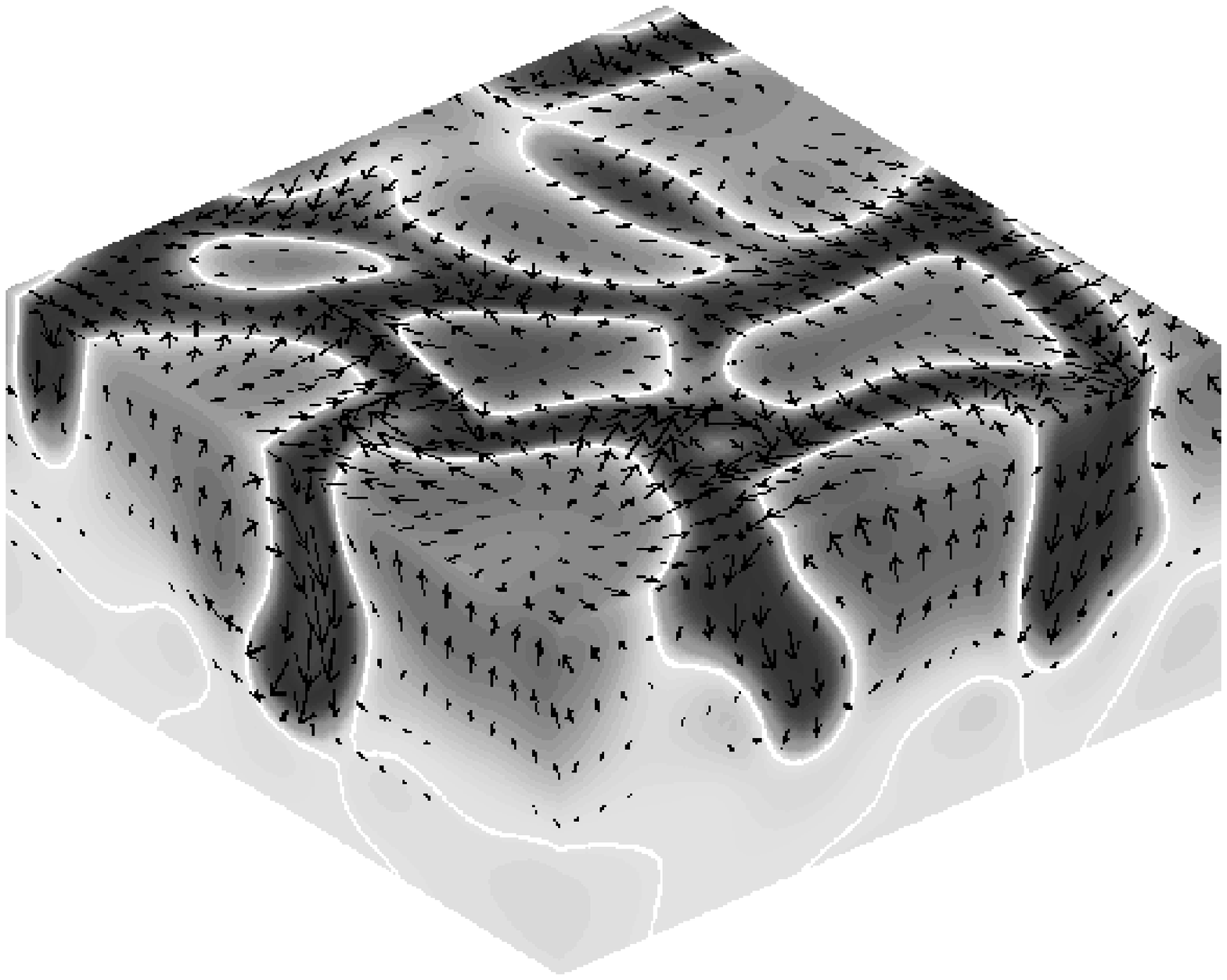,width=9cm}}
\centerline{\psfig{file=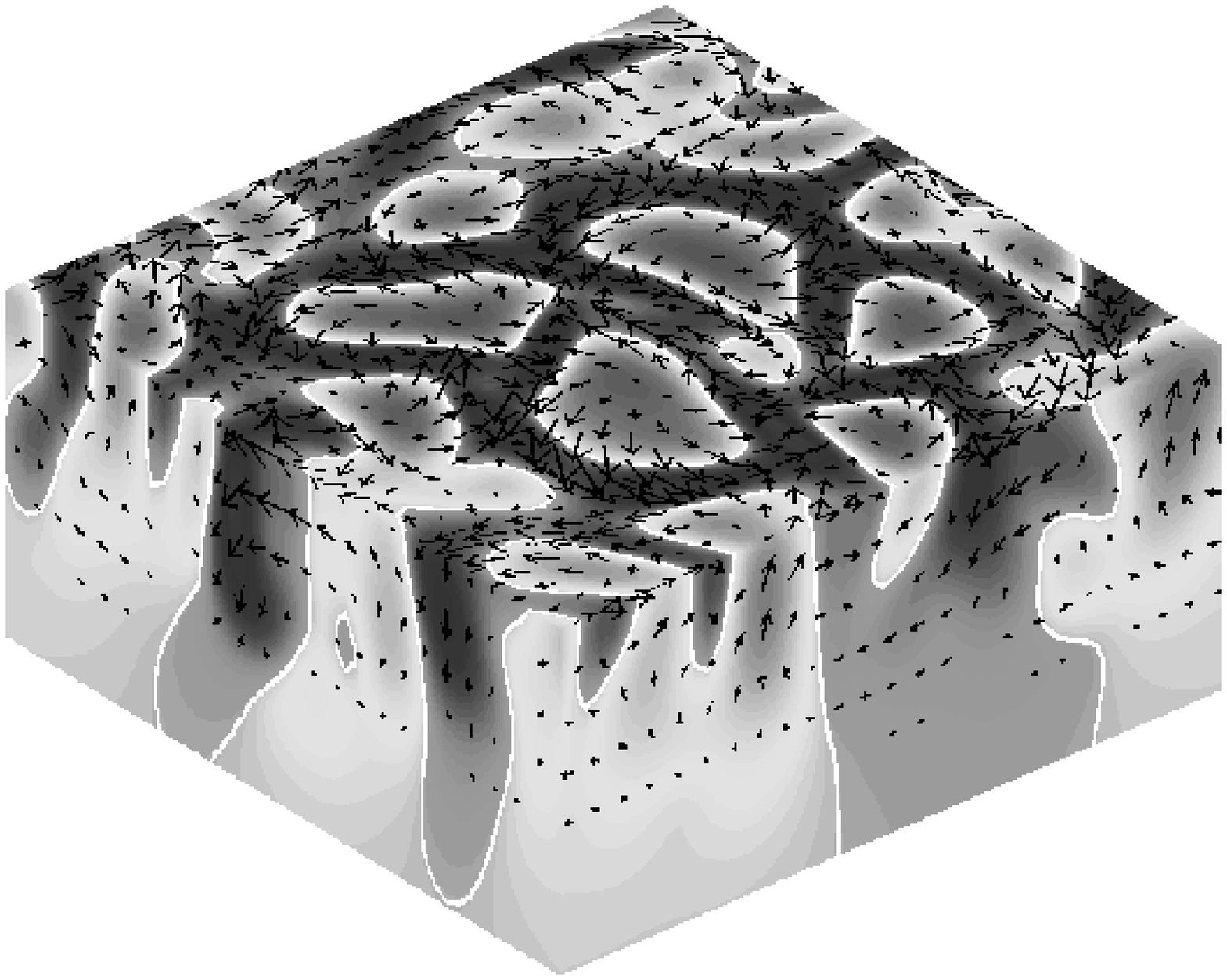,width=9cm}}
\caption{Snapshot of the vertical velocity (connected lanes: downward,
         isolated  patches: upward). {\em Top}\/: run 6, moderate rotation. 
         {\em Bottom}\/: run 9, strong rotation. The white curve denotes
         the zero level of the vertical velocity. The top and bottom
         surfaces correspond to $z=0.09$ and $z=1.81$, respectively.}
                                                             \label{fig7}
\end{figure} 

In order to shed further light on the influence of rotation, Fig.~\ref{fig7} shows snapshots of 
the vertical velocity for two simulations with different rotation rates. 
Clearly, stronger rotation causes the eddies to have a smaller diameter in the horizontal plane 
(i.e. perpendicular to the rotation axis). Hence the number of eddies within the simulation domain 
increases with the rotation rate, and this leads to a more accurate result for $\alpha$, as is obvious
from the length of the error bars in Fig. \ref{pta}, taking into account the duration of the 
runs (Table \ref{paramtable}). In the simpler case of Boussinesq convection in a rotating layer, 
the scale at which the instability sets in also decreases with increasing Taylor number 
(Chandrasekhar \cite{chandra61}). Although the present simulations are done at supercritical Rayleigh 
numbers whereas the analysis of Chandrasekhar applies to the marginally stable case, we may
for the moment set aside this difference, as well as possible effects of compressibility and 
different boundary conditions. In fact, the smallest value of the local Rayleigh number in the 
unstable layer was found to be typically $10^4$, so that the actual supercriticality is less than assumed.
The present runs, which are characterized by $\mb{Pr}=\mb{Pm}=1$ and $\mb{Ch}=0.018$, belong to a regime 
where the onset of instability occurs in the form of non-oscillatory convection. 
Since the value of the critical Rayleigh number depends on the horizontal wavenumber, $k$, there
is a critical wavenumber, $k_{\rm c}$, corresponding to a preferred eddy size $2\pi/k_{\rm c}$, 
for which the critical Rayleigh number attains a minimal value, $\mb{Ra}_{\rm c}$. 
The eddy sizes observed in the simulations are in quite good agreement with the analytical
values of $2\pi/k_{\rm c}$ for convection between two free boundary surfaces
(Table \ref{tablesize}). The results suggest that the preferred mode of convection 
is the same as that pertaining to the onset of convection.
The critical Rayleigh number increases with $\mb{Ta}$ from $661$ for $\mb{Ta}=2$ up to
$2.13\cd 10^4$ for $\mb{Ta}=10^5$. This explains why for a fixed 
value of $\mb{Ra}$ the intensity of convection, as measured by $u_{\rm rms}$, decreases with 
increasing rotation rate. Since the actual Rayleigh number in the unstable layer has a local 
minimum of about $10^4$, this also explains why no convection was observed for $\mb{Ta}=10^5$.

\begin{figure}[htb]
\centerline{\psfig{file=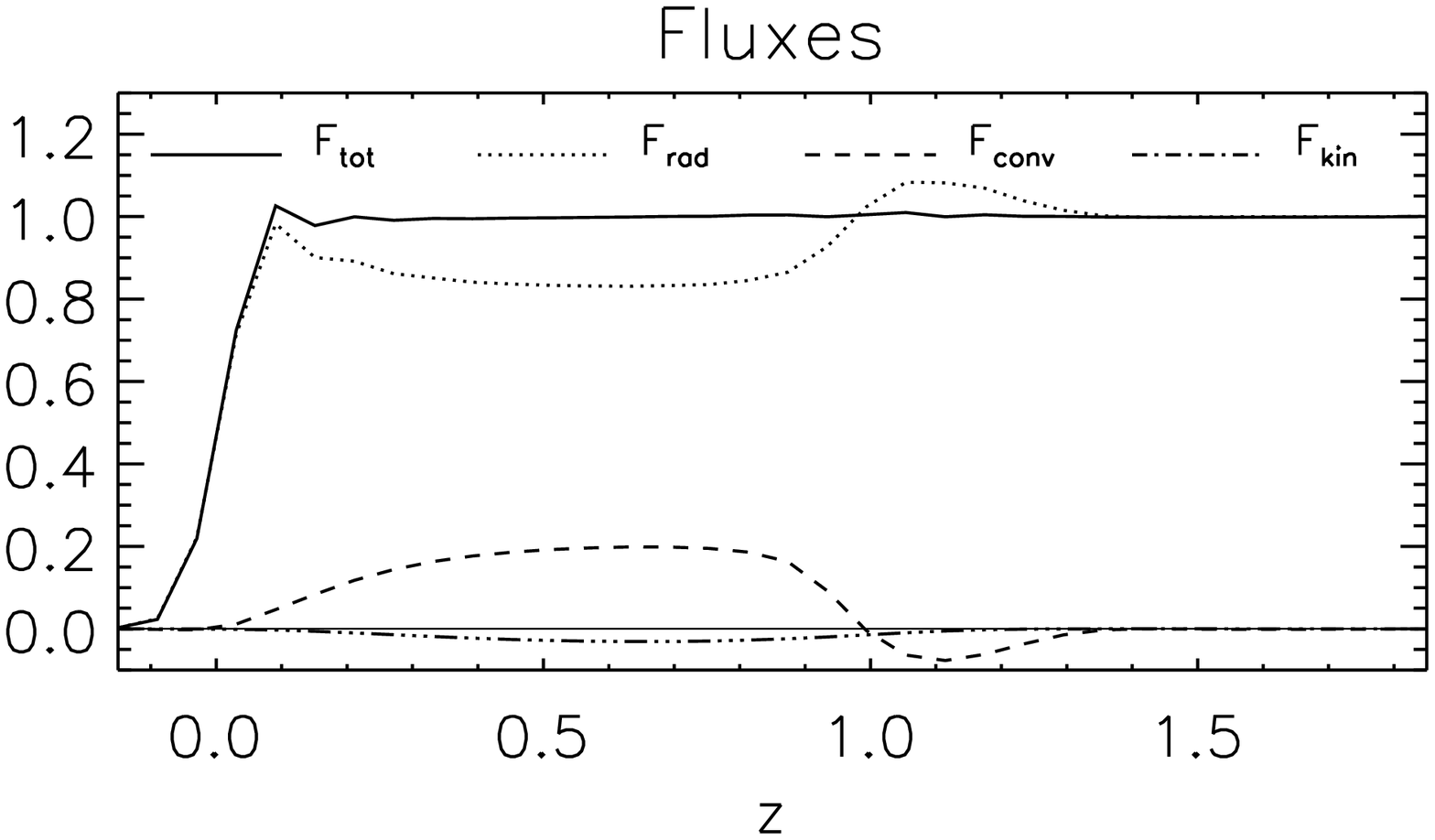,width=9cm}}
\centerline{\psfig{file=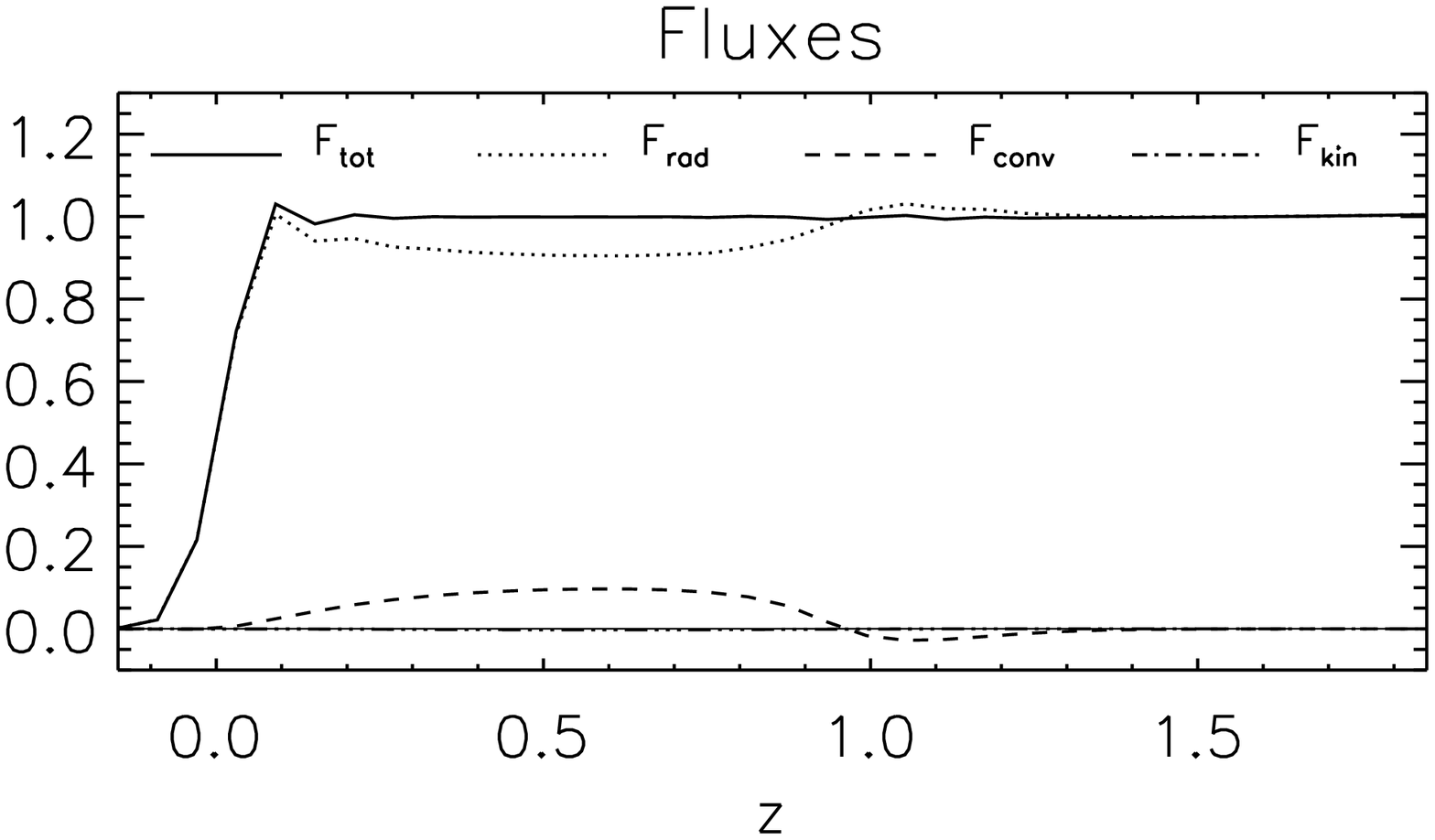,width=9cm}}
\caption{Energy fluxes. {\em Top}\/: run 6; {\em Bottom}\/: run 9.
         Only the three most important contributions are shown, the viscous
         and electromagnetic fluxes being negligible. The total flux
         $\vec{F}_{\rm tot}$ does not include the flux associated with the
         cooling term $Q$, Eq. (\protect{\ref{coolingterm}}); 
         hence the steep rise in region 1.}                  \label{figflux}
\end{figure}

As the scale of the individual eddies decreases, the corresponding
decrease of the effective Coriolis number, $2\Omega \ell/u_{\rms}$, may
be interpreted as a readjustment of the convection in response to
increased Coriolis forces, which tend to hamper the convective energy
transport. Still the convective flux of the run with the strongest
rotation (run 9) is reduced by a factor of about two compared to runs
with weak rotation. Apart from a possible effect of the increased
Coriolis force, the reduced flux may result from the fact that smaller
eddies lead to larger temperature gradients in horizontal planes, and this
causes increased horizontal radiative transport between adjacent eddies
which reduces the efficiency of convective energy transport in the 
vertical direction (Fig.~\ref{figflux}).

\section{Alpha and magnetic field strength}

If the strength of the imposed magnetic field is increased, effects of
the Lorentz force become more noticeable. Moreover, these effects depend
on the orientation of the imposed field, i.e. vertical or horizontal. 
From the previously discussed runs, a representative case (run 7) 
was selected as the starting point for runs with increasingly 
strong imposed magnetic fields of both orientations (Table \ref{table3}). 
A comprehensive study of the mode structure of magnetoconvection for 
different imposed magnetic fields is beyond the scope of the present paper,
though. Various cases for a vertical imposed field, without rotation, are
treated by Matthews et al. (\cite{matthews95}). Wissink et al. (2000)
consider the breakup of a horizontal magnetic layer into a number of
flux tubes, thereby including the effect of rotation.

\begin{figure}
\centerline{\psfig{file=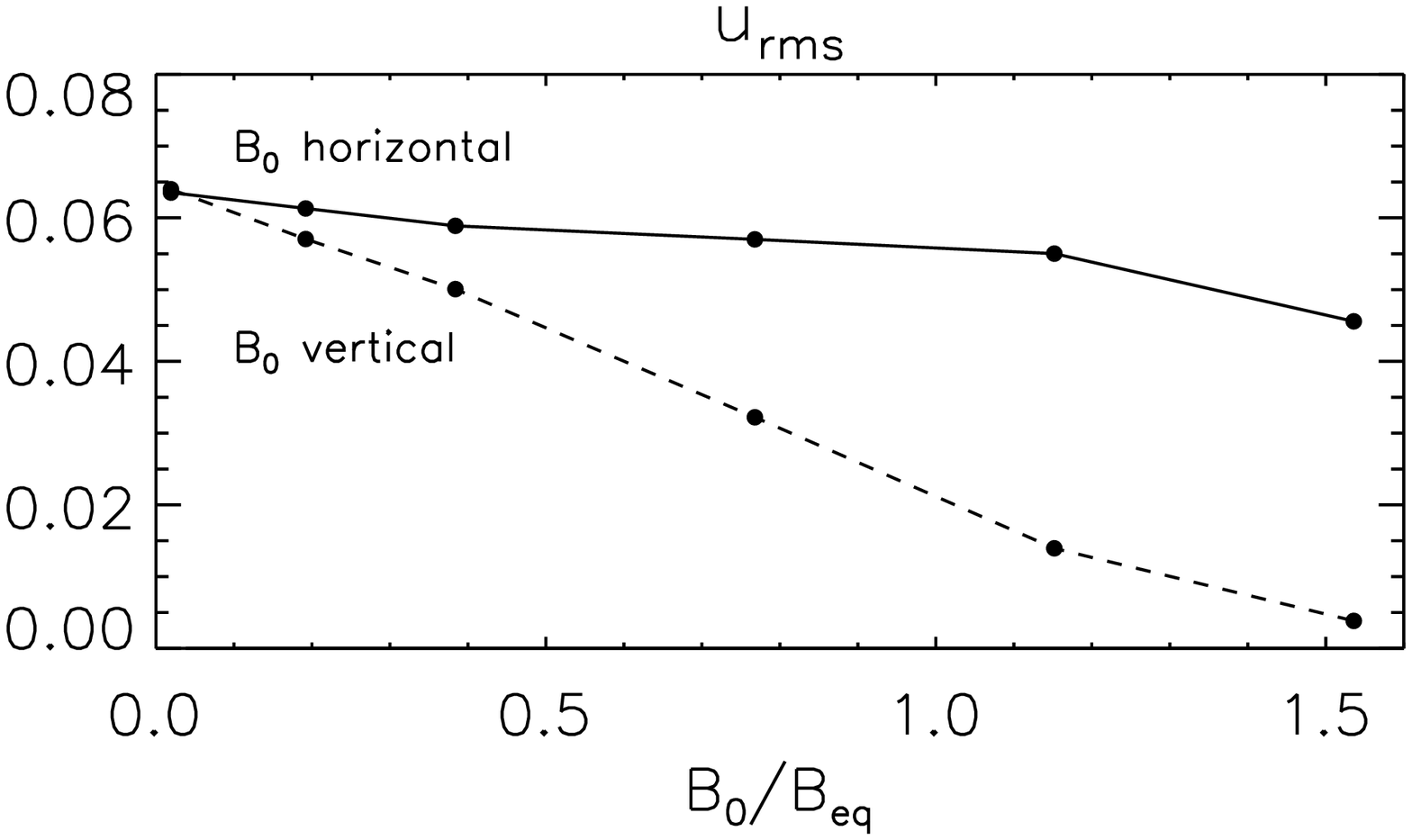,width=8.7cm}}
\centerline{\psfig{file=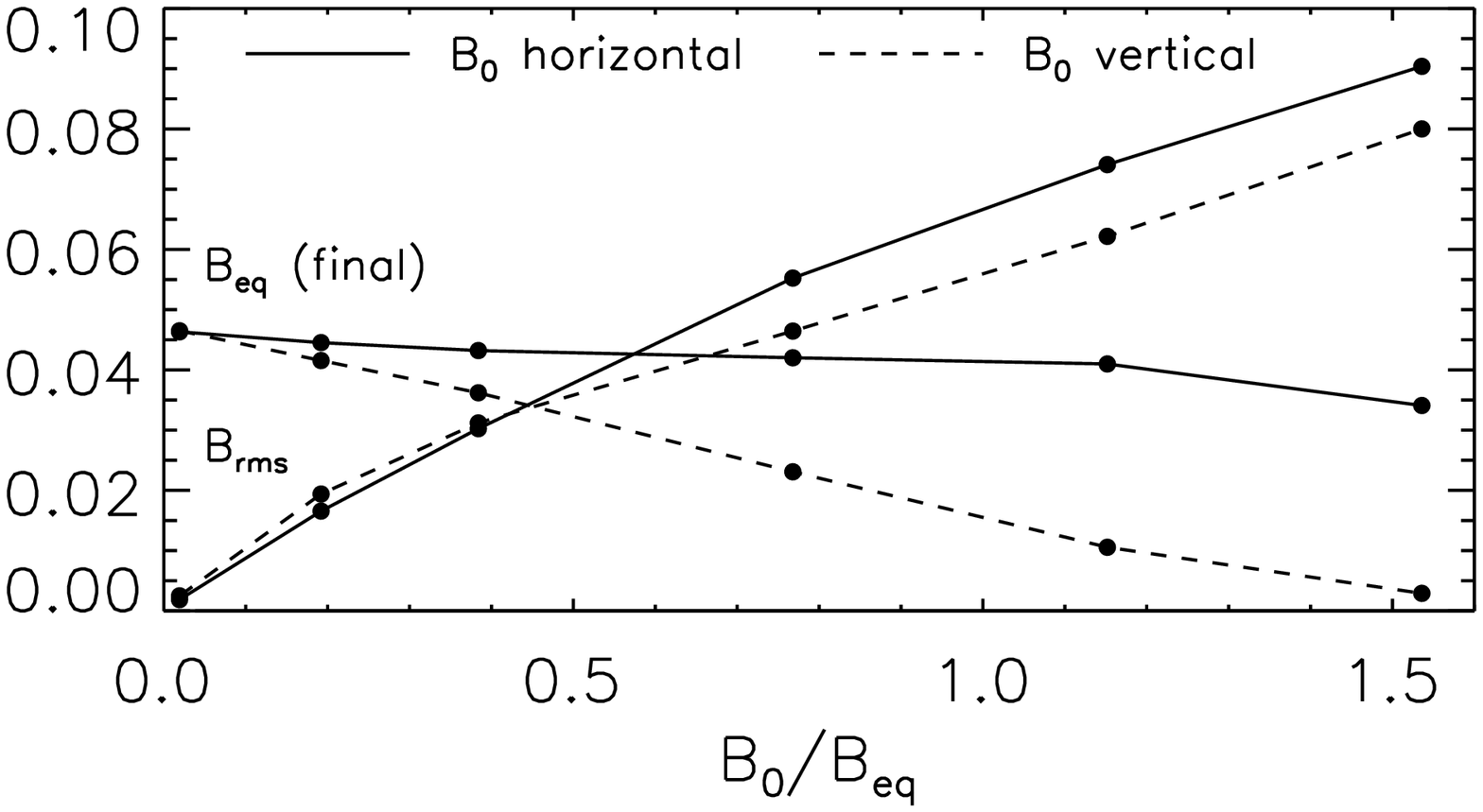,width=8.7cm}}
\caption{{\em Top}\/: rms velocity perturbations as a function of the 
         strength of the imposed magnetic field (normalized to the value
         of $B_{\rm eq}$ for small $B_0$). {\em Bottom}\/: rms magnetic
         field and final equipartition field strength.}     \label{purmsb0}
\end{figure} 

\begin{table}[htb]
\caption{Initial field strength, $B_0$, and averaging interval, $T$. Other 
         parameters as in run 7 (Table \protect{\ref{paramtable}}). The 
         imposed magnetic field is $\vec{B}_0=B_0\vec{e}_z$, or $B_0\vec{e}_x$,
         depending on whether $\alpha_{\rm V}$ or $\alpha_{\rm H}$ 
         is determined.} \label{table3}
\begin{center}
\begin{tabular}{ccccccc} \hline\noalign{\smallskip}
run                    & 7       & 7a     & 7b    & 7c    & 7d      & 7e   \\
\noalign{\smallskip}\hline\noalign{\smallskip}
$B_0$                  & $0.001$ & $0.01$ & 0.02  & 0.04  & $0.06$  & 0.08 \\
$T$ ($\alpha_{\rm V}$) & $940$   & $205$  & 203   & 200   & $414$   & 96   \\
$T$ ($\alpha_{\rm H}$) & $508$   & $280$  & 278   & 559   & $381$   & 268  \\
\noalign{\smallskip}\hline
\end{tabular}
\end{center}
\end{table}

\begin{figure}[htb]
\centerline{\psfig{file=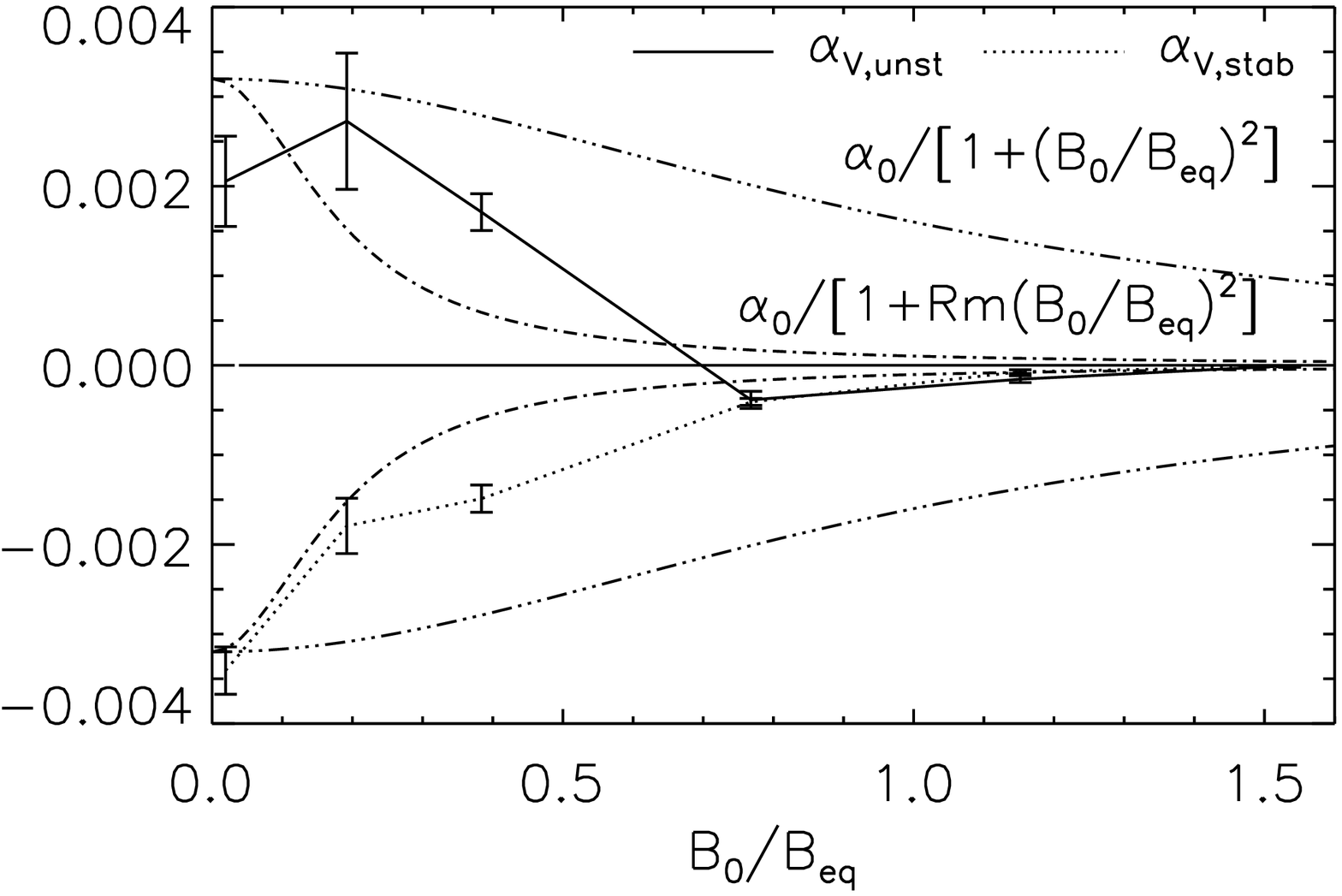,width=8.5cm}}
\centerline{\psfig{file=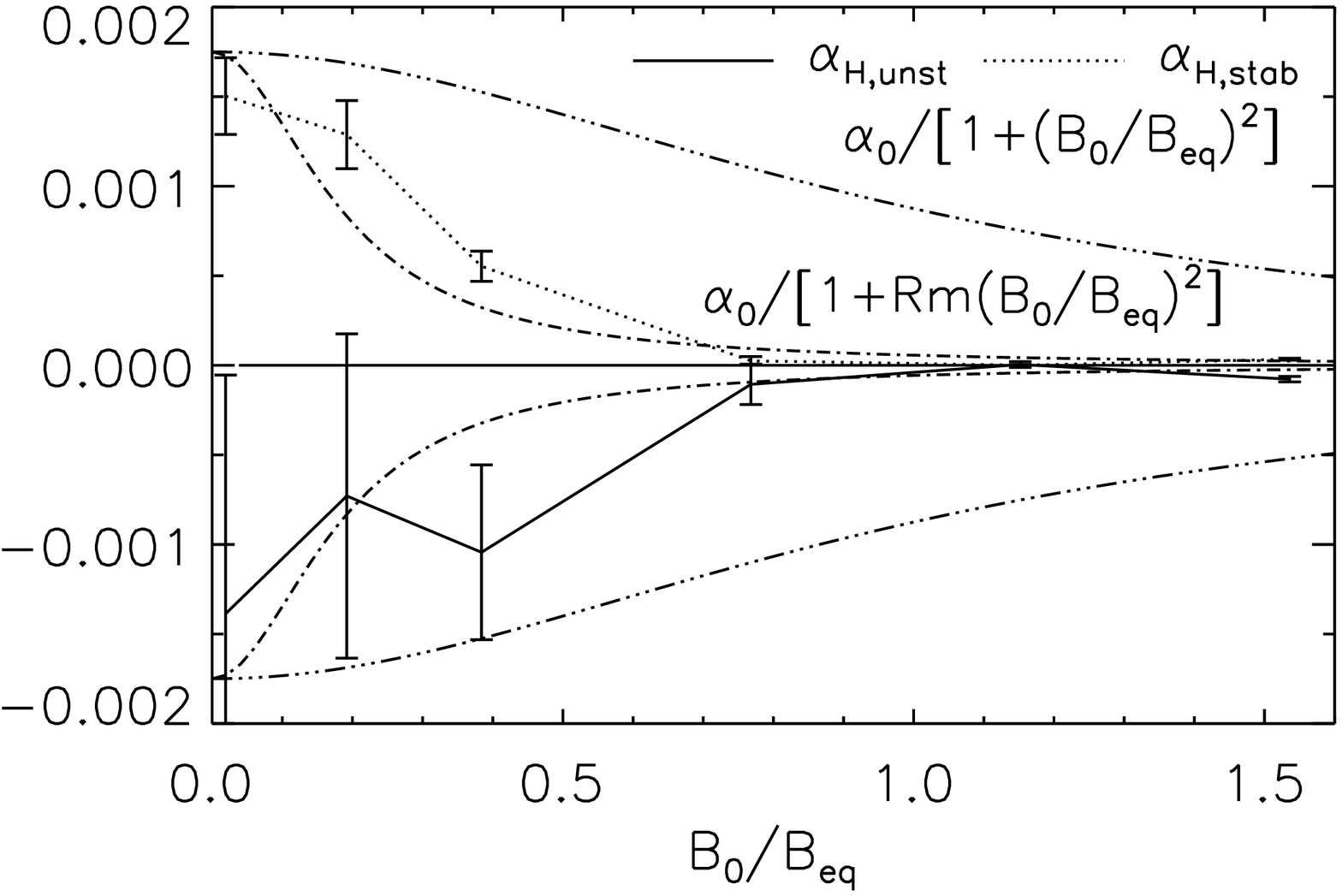,width=8.5cm}}
\caption{{\em Top}\/: $\alpha_{\rm V,unst}$ and $\alpha_{\rm V,stab}$ as
         functions of $B_0/B_{\rm eq}$ for run 7, where 
         $\vec{B}_0=B_0\vec{e}_z$. {\em Bottom}\/: $\alpha_{\rm H,unst}$ 
         and $\alpha_{\rm H,stab}$ as functions of $B_0/B_{\rm eq}$ for 
         run 7, where $\vec{B}_0=B_0\vec{e}_x$. Here $B_{\rm eq}=0.052$ is
         the equipartition field strength, averaged over the unstable layer,
         for convection without a magnetic field. Analytical quenching
         results ({\em dash-dotted}) are given for both signs. All 
         calculations are for the solar south pole.}        \label{figpb0}
\end{figure}

For an imposed magnetic field in the vertical direction, convection is 
increasingly hampered
with increasing field strength, as is indicated by a decrease of the rms turbulent velocity 
(Fig. \ref{purmsb0}, top) and of the convective energy flux. Since the magnetic energy is enhanced 
by advection of the imposed field, while the kinetic energy decreases as a result of the Lorentz 
force, after a while the rms magnetic field can be much larger than the 
equipartition field strength (Fig. \ref{purmsb0}, bottom). 
Moreover, if the initial magnetic field is too strong, then convection
dies out completely. This in fact occured during
run 7e ($B_0/B_{\rm eq}\approx 1.7$). After 100 time units, the
volume-averaged energy density of the velocity and magnetic-field 
perturbations had both fallen by about $2$ orders of magnitude.
The decrease of $\alpha_{\rm V}$ is more rapid than that of $u_{\rm rms}$, 
because advection of the imposed field by convection is stifled; this 
explains why alpha decreases also when measured in dynamical units 
(i.e., when divided by $u_{\rms}$).

A debate is ongoing in the literature about how and whether the magnetic
Reynolds number affects the magnetic quenching of the $\alpha$ effect. 
The dependence of $\alpha$ on the magnetic field strength is often 
schematically represented in the form 
$\alpha\approx\alpha_0[1+\mb{Rm}^p B^2/B_{\rm eq}^2]^{-1}$. 
While some numerical and analytical results suggest that $p\approx 0$
(Kraichnan \cite{kraichnan79a}, \cite{kraichnan79b}, 
Brandenburg \& Donner \cite{brandenburg97}), 
others suggest $p\approx 1$ (Cattaneo \& Vainshtein 
\cite{cattaneo91b}, Vainshtein \& Cattaneo \cite{vainshtein92}, 
Tao et al. \cite{tao93}, Cattaneo \cite{cattaneo94}, Cattaneo \& Hughes 1996,
Vainshtein \cite{vainshtein98}, Brandenburg \cite{brandenburg00}).
Although the present set of simulations is not well-suited to answer this
question because of the relatively small Reynolds numbers 
($\mb{Rm}\approx 30$), the results for both $\alpha_{\rm V}$ and
$\alpha_{\rm H}$ are consistent with $p\approx 1$ (Fig.~\ref{figpb0}). 
Blackman \& Field (\cite{blackman00}) argued that the quenching with 
$p\approx 1$ observed in simulations with periodic boundaries may in
fact be a consequence of the boundary conditions rather than a dynamic
effect, and that $p=0$ should be possible if a flux of magnetic helicity
through the
boundaries is allowed. Our conditions (\ref{boundy}) do allow for a flux
of magnetic helicity through the top and bottom boundaries. Thus, the fact
that $\alpha$ quenching sets in when $B^2\sim\mbox{Rm}B_{\rm eq}^2$
suggests that an Rm-dependence of the $\alpha$ quenching is possible
in the present case; such quenching could be dynamic according to the 
reasoning of Blackman and Field. However, this conclusion may change if one
allows for shear near the surface. This is the case between the solar surface
and the corona, and so a strong helicity flux is possible in that case
(Berger \& Ruzmaikin \cite{berger00}).

It appears that for boundaries that allow for a flux of helicity
diverse Rm dependencies of the $\alpha$ quenching are possible,
depending, e.g., on the character of the forcing of the flow.
Brandenburg and Dobler (2001) report a case where $p\approx 1/2$.
We also note that even with the strong $p\approx 1$-quenching of
$\alpha$ a large-scale field may build up because of an equally strong
quenching of the turbulent magnetic diffusivity (Brandenburg 
\cite{brandenburg00}).

The magnetic-field dependence of the $\alpha$ coefficients is essentially
the same for the cases of the vertical and horizontal imposed fields,
even though convection is hampered less in the case of a horizontal
imposed field (Fig. \ref{figpb0}, top). For field strengths comparable to
the equipartition value, the convective pattern becomes very different. If 
$B_0=0.02\approx 0.38B_{\rm eq}$, convection exhibits a pattern of
irregular oblique elongated cells that make an angle with the direction
of the imposed magnetic field, and a small horizontal $\alpha$ effect 
is still measured. For runs with field strengths 
$B_0=0.04\approx 0.77 B_{\rm eq}$ and $B_0=0.06\approx 1.15 B_{\rm eq}$
respectively, longitudinal rolls are formed that are increasingly aligned
along the $x$-direction, and the coefficient $\alpha_{\rm H}$ is 
practically zero. For still stronger fields 
($B_0=0.08\approx 1.5B_{\rm eq}$), convection assumes the form of highly
regular oblique lanes, and a small $\alpha$ effect is again measured.
 
\section{Discussion}
The following discussion of our results in the solar context may be considered
as complimentary to that of Brandenburg et al. (\cite{brandenburg90}).
First-order smoothing is a justified approximation if either the magnetic
Reynolds number is small, or if the flow is slow in the sense $u\tau/d\ll 1$.
In our dimensionless variables, the latter condition would mean
$u_{\rms}\tau/d \ll 1$ but, with the rms velocities and coherence times 
obtained in our simulations, we rather have $u_{\rms}\tau/d \ga 1$. 
Our magnetic Reynolds
number (20--35) is quite moderate but not small. Hence the simulations clearly
go beyond first-order smoothing, although we are far from the conditions met
in the solar convection zone. The large fluctuations that occur at large
Reynolds number are evident in our calculations. In particular we see that
the components of the $\alpha$ tensor undergo large fluctuations. Only
after averaging over many coherence times and over the horizontal coordinates
of the box a significant depth variation is obtained. With this variation
we confirm some results obtained in the case of weakly anisotropic
turbulence, and extend them beyond the limits of first-order smoothing:
in the southern hemisphere
the horizontal $\alpha$ coefficient (Fig. \ref{pkinhel}, row 2) is
negative (positive) in the upper (lower) part of the box, as is plausible
from the rotational effect on convergent and divergent velocities that arise
as the flow is forced into the horizontal direction. On the other hand we
find the opposite sign for the vertical $\alpha$ coefficient (Fig. \ref{fig2}
and Fig. \ref{pkinhel}, row 1). Also of opposite sign is the small-scale
current helicity. Indeed, both the sign and the rotational dependence of
$\alpha_{\rm V}$ in the unstable layer roughly agree with that predicted
by the simple FOSA-expression which relates $\alpha$ to the current 
helicity (\ref{alp2}).

For cases with convective overshooting in the lower part of the box the
components of $\alpha$ reverse their sign near the transition to the
stable stratification. For an $\alpha\Omega$ dynamo the horizontal component
$\alpha_{\rm H}$ is the essential ingredient; possibly the sign change of
$\alpha_{\rm H}$ has interesting implications for a dynamo that operates
as an interface wave at the base of the convection zone. In the layer of
convective overshooting the sign of $\alpha_{\rm H}$ is appropriate for
the equatorward migration of the mean field, if the positive shear 
$\partial\Omega/\partial r$, as inferred from helioseismology for the depth
range around $0.7r_\odot$ and the lower heliographic latitudes (Kosovichev
et al. \cite{kosovichev97}), is used as the other essential ingredient. 
These conclusions are drawn from simulations with an imposed magnetic 
field $B_0=10^{-3}$, and may still be valid for $B_0=10^{-2}$, as the
results of run 7a, shown in Fig. \ref{figpb0}, suggest. The latter field
strength corresponds to $\approx 10\,$T, which is the strength of the
field in the stable layer below the convection zone, as deduced from
the properties of rising flux tubes (Caligari et al. \cite{caligari95}). 
For stronger fields we obtain a marked quenching of $\alpha$ 
consistent with a factor $[1+\mb{Rm}\,(B_0/B_{\rm eq})^2]^{-1}$, although
we have checked this only for $\mb{Rm} \approx 30$. 
The simulations have also revealed a striking difference between the
convective patterns for a vertical and a horizontal orientation of the
imposed magnetic field, with rolls being formed for strong horizontal fields.
 
In view of a strong toroidal field at the base of the convection zone a 
magnetic Rayleigh-Taylor instability 
(Brandenburg \& Schmitt \cite{brandenburg98}) or an instability 
of toroidal flux tubes (Ferriz-Mas et al. \cite{ferrizmas94})
should be employed to obtain an $\alpha$ effect. Nevertheless, since the sign
of $\alpha_{\rm H}$ follows plausible arguments, we may hope to obtain the
same sign for such a dynamically determined $\alpha$
(Brandenburg \& Schmitt \cite{brandenburg98}).

As far as the dependence of $\alpha$ on rotation is concerned, the simulation
may yield more reliable results than for the other parameters. In the solar
convection zone the Coriolis number is small near the surface, and increases to
values around 1 or somewhat larger near its base. This is the regime shown
in Fig. \ref{pta}. The rotational quenching of $\alpha_{\rm V}$ and the
saturation of $\alpha_{\rm H}$ (Fig. \ref{pta}, top) may therefore just
begin in the depth region where the dynamo operates, but in solar-type stars
which are younger and therefore rotating faster, these effects may be 
important. We have observed some agreement between the numerical results for
$\alpha$ and the predictions of R\"udiger \& Kichatinov (\cite{ruediger93}),
as far as the dependence on rotation and on depth is concerned.
In both the unstable and the stable layer, the approximate Coriolis number
dependence and the correct sign for $\alpha_{\rm H}$ are reproduced. The
main discrepancies concern the amplitude of $\alpha$ (the scaling factors 
of the theoretical curves in Fig.~\ref{pta}) and the sign of
$\alpha_{\rm V}$ in the unstable layer, as well as the sign of the 
small-scale current helicity.

Even within the limited range accessible to numerical simulation, the 
parameter space is enormous. We have not yet considered the latitude
dependence of $\alpha$ (although some earlier work of Brandenburg 
(\cite{brandenburg94}) indicates that $|\alpha|$ is maximum at about
$\pm 60^\circ$ of latitude), and we have not yet explored the dependencies
on the Reynolds and Rayleigh numbers. 
This must be done by further simulations.

\end{document}